\documentclass[natbib]{nature_fig}
\usepackage{amssymb}
\usepackage{amsmath}
\usepackage{txfonts}
\usepackage{threeparttable}

\usepackage{hyperref}
\hypersetup{
colorlinks=true,
linkcolor=blue,
citecolor=blue,
urlcolor=blue
}

\usepackage{graphicx}
\usepackage{caption}
\usepackage{url}
\usepackage{subcaption}

\usepackage{tikz}
\usetikzlibrary{shapes.geometric, arrows}

\usepackage{lscape}
\usepackage{longtable}
\usepackage{multirow} 
\usepackage{threeparttable}
\usepackage{threeparttablex}
\usepackage{pdflscape}

\newcommand{\apjl}{Astrophys. J. Lett.}   
\newcommand{\apjs}{Astrophys. J. Suppl. Ser.}   
\newcommand{\ao}{Appl. Opt.}   
\newcommand{\aap}{Astron. Astrophys.}   

\newcommand{\efluxunit}{\rm erg\,s^{-1}\, cm^{-2}}
\newcommand{\fluxunit}{\rm ph\,s^{-1}\, cm^{-2}}
\newcommand{\emin}{E_{\rm min}\,=\, 100\, \rm MeV}
\newcommand{\emid}{E_{\rm min}\,=\, 1\, \rm GeV}
\newcommand{\emax}{E_{\rm max}\,=\, 300\, \rm GeV}
\newcommand{\LnuI}{L_{\rm 1.4\,GHz}}
\newcommand{\LnuIII}{L_{\rm 3\,GHz}}
\newcommand{\LnuV}{L_{\rm 5\,GHz}}
\newcommand{\LIR}{L_{\rm 8-1000 \,\mu m}}
\newcommand{\Lsoftx}{L_{\rm 0.3-10\,keV}}
\newcommand{\fx}{f_{\rm 14-195\,keV}}
\newcommand{\Lx}{L_{\rm 14-195\,keV}}
\newcommand{\lamX}{\lambda_{\rm X}}
\newcommand{\Rrx}{\mathcal{R}_{\rm rX}}
\newcommand{\Rrb}{\mathcal{R}_{\rm rB}}

\newcommand{\Ledd}{L_{\rm Edd}}

\newcommand{\LGam}{L_{\rm 1-300\,GeV}}
\newcommand{\Lsun}{L_{\odot}}
\newcommand{\Mbh}{M_{\bullet}}
\newcommand{\Msun}{M_{\odot}}
\newcommand{\Rg}{R_{\rm g}}
\newcommand{\Rcom}{R_{\rm com}} 
\newcommand{\taugg}{\tau_{\gamma\gamma}(\nu)}
\newcommand{\sigmagg}{\sigma_{\gamma\gamma}}
\newcommand{\sigmaT}{\sigma_{\rm T}}
\newcommand{\xinth}{\xi_{\rm nth}}
\newcommand{\xicom}{\xi_{\rm com}}
\newcommand{\xiext}{\xi_{\rm ext}}
\newcommand{\Rext}{R_{\rm ext}}

\newcommand{\Lnup}{L_\nu^\prime}
\newcommand{\Next}{n_{\rm ext}}
\newcommand{\Ncom}{n_{\rm com}}
\newcommand{\ergs}{\rm erg\,s^{-1}}

\usepackage[utf8]{inputenc}

\title{Fermi detection of gamma-ray emission from the hot coronae of radio-quiet active galactic nuclei}

\author{
\centering
Jun-Rong~Liu$^{1,2}$\href{https://orcid.org/0000-0003-3086-7804}{\includegraphics[scale=0.08]{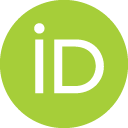}},
Jian-Min~Wang$^{1,2,3\dagger}$\href{https://orcid.org/0000-0001-9449-9268}{\includegraphics[scale=0.08]{ORCIDiD.png}},\\
and 
S.~Abdollahi$^{4}$,
M.~Ajello$^{5}$,
R.~Alves~Batista$^{6}$,
L.~Baldini$^{7}$,
C.~Bartolini$^{8,9}$,
D.~Bastieri$^{10,11,12}$,
J.~Becerra~Gonzalez$^{13}$,
R.~Bellazzini$^{14}$,
B.~Berenji$^{15}$,
E.~Bissaldi$^{16,8}$,
R.~D.~Blandford$^{17}$,
R.~Bonino$^{18,19}$,
P.~Bruel$^{20}$,
S.~Buson$^{21}$,
R.~A.~Cameron$^{17}$,
P.~A.~Caraveo$^{22}$,
E.~Cavazzuti$^{23}$,
G.~Chiaro$^{22}$,
N.~Cibrario$^{18,19}$,
S.~Ciprini$^{24,25}$,
P.~Cristarella~Orestano$^{26,27}$,
S.~Cutini$^{27}$,
F.~D'Ammando$^{28}$,
N.~Di~Lalla$^{17}$,
A.~Dinesh$^{29}$,
L.~Di~Venere$^{8}$,
A.~Dom\'inguez$^{29}$,
S.~J.~Fegan$^{20}$,
A.~Fiori$^{7}$,
A.~Franckowiak$^{30}$,
Y.~Fukazawa$^{31}$,
S.~Funk$^{32}$,
P.~Fusco$^{16,8}$,
F.~Gargano$^{8}$,
C.~Gasbarra$^{24,33}$,
D.~Gasparrini$^{24,25}$,
S.~Germani$^{26,27}$,
N.~Giglietto$^{16,8}$,
M.~Giliberti$^{8,16}$,
F.~Giordano$^{16,8}$,
M.~Giroletti$^{28}$,
D.~Green$^{34}$,
I.~A.~Grenier$^{35}$,
S.~Guiriec$^{36,37}$,
M.~Hashizume$^{31}$,
E.~Hays$^{37}$,
J.W.~Hewitt$^{38}$,
D.~Horan$^{20}$,
Xian~Hou$^{39,40}$,
C.~Karwin$^{37}$,
T.~Kayanoki$^{31}$,
M.~Kuss$^{14}$,
A.~Laviron$^{20}$,
M.~Lemoine-Goumard$^{41}$,
Jian~Li$^{42,43\dagger}$\href{https://orcid.org/0000-0003-1720-9727}{\includegraphics[scale=0.08]{ORCIDiD.png}},
I.~Liodakis$^{44}$,
F.~Longo$^{45,46}$,
F.~Loparco$^{16,8}$,
L.~Lorusso$^{16,8}$,
P.~Lubrano$^{27}$,
S.~Maldera$^{18}$,
L.~Marcotulli$^{47,5}$,
G.~Mart\'i-Devesa$^{45}$,
M.~N.~Mazziotta$^{8}$,
I.Mereu$^{27,26}$,
P.~F.~Michelson$^{17}$,
N.~Mirabal$^{37,48}$,
W.~Mitthumsiri$^{49}$,
T.~Mizuno$^{50}$,
M.~E.~Monzani$^{17,51}$,
T.~Morishita$^{31}$,
A.~Morselli$^{24}$,
I.~V.~Moskalenko$^{17}$,
M.~Negro$^{52}$,
R.~Niwa$^{31}$,
N.~Omodei$^{17}$,
M.~Orienti$^{28}$,
E.~Orlando$^{53,17}$,
J.~F.~Ormes$^{54}$,
D.~Paneque$^{34}$,
G.~Panzarini$^{16,8}$,
M.~Persic$^{46,55}$,
M.~Pesce-Rollins$^{14}$,
R.~Pillera$^{16,8}$,
T.~A.~Porter$^{17}$,
G.~Principe$^{45,46,28}$,
S.~Rain\`o$^{16,8}$,
R.~Rando$^{11,10,12}$,
B.~Rani$^{37,48}$,
M.~Razzano$^{7}$,
A.~Reimer$^{56}$,
O.~Reimer$^{56}$,
M.~S\'anchez-Conde$^{6,57}$,
P.~M.~Saz~Parkinson$^{58}$,
D.~Serini$^{8}$,
C.~Sgr\`o$^{14}$,
E.~J.~Siskind$^{59}$,
G.~Spandre$^{14}$,
P.~Spinelli$^{16,8}$,
D.~J.~Suson$^{60}$,
H.~Tajima$^{61,62}$,
J.~B.~Thayer$^{17}$,
D.~F.~Torres$^{63,64}$,
Zi-Hao~Zhao$^{43}$\\
(Fermi-LAT Collaboration)
\begin{enumerate}
\item[1.]~Key Laboratory for Particle Astrophysics, Institute of High Energy Physics, Beijing 100049, China
\item[2.]~School of Astronomy and Space Sciences, University of Chinese Academy of Sciences, Beijing 100049, China
\item[3.]~National Astronomical Observatory of China, Beijing 100020, China
\item[4.]~IRAP, Universit\'e de Toulouse, CNRS, UPS, CNES, F-31028 Toulouse, France
\item[5.]~Department of Physics and Astronomy, Clemson University, Kinard Lab of Physics, Clemson, SC 29634-0978, USA
\item[6.]~Instituto de F\'isica Te\'orica UAM/CSIC, Universidad Aut\'onoma de Madrid, E-28049 Madrid, Spain
\item[7.]~Universit\`a di Pisa and Istituto Nazionale di Fisica Nucleare, Sezione di Pisa I-56127 Pisa, Italy
\item[8.]~Istituto Nazionale di Fisica Nucleare, Sezione di Bari, I-70126 Bari, Italy
\item[9.]~Universit\`a degli studi di Trento, via Calepina 14, 38122 Trento, Italy
\item[10.]~Istituto Nazionale di Fisica Nucleare, Sezione di Padova, I-35131 Padova, Italy
\item[11.]~Dipartimento di Fisica e Astronomia ``G. Galilei'', Universit\`a di Padova, Via F. Marzolo, 8, I-35131 Padova, Italy
\item[12.]~Center for Space Studies and Activities ``G. Colombo", University of Padova, Via Venezia 15, I-35131 Padova, Italy
\item[13.]~Instituto de Astrof\'isica de Canarias and Universidad de La Laguna, Dpto. Astrof\'isica, 38200 La Laguna, Tenerife, Spain
\item[14.]~Istituto Nazionale di Fisica Nucleare, Sezione di Pisa, I-56127 Pisa, Italy
\item[15.]~California State University, Los Angeles, Department of Physics and Astronomy, Los Angeles, CA 90032, USA
\item[16.]~Dipartimento di Fisica ``M. Merlin" dell'Universit\`a e del Politecnico di Bari, via Amendola 173, I-70126 Bari, Italy
\item[17.]~W. W. Hansen Experimental Physics Laboratory, Kavli Institute for Particle Astrophysics and Cosmology, Department of Physics and SLAC National Accelerator Laboratory, Stanford University, Stanford, CA 94305, USA
\item[18.]~Istituto Nazionale di Fisica Nucleare, Sezione di Torino, I-10125 Torino, Italy
\item[19.]~Dipartimento di Fisica, Universit\`a degli Studi di Torino, I-10125 Torino, Italy
\item[20.]~Laboratoire Leprince-Ringuet, CNRS/IN2P3, \'Ecole polytechnique, Institut Polytechnique de Paris, 91120 Palaiseau, France
\item[21.]~Institut f\"ur Theoretische Physik and Astrophysik, Universit\"at W\"urzburg, D-97074 W\"urzburg, Germany
\item[22.]~INAF-Istituto di Astrofisica Spaziale e Fisica Cosmica Milano, via E. Bassini 15, I-20133 Milano, Italy
\item[23.]~Italian Space Agency, Via del Politecnico snc, 00133 Roma, Italy
\item[24.]~Istituto Nazionale di Fisica Nucleare, Sezione di Roma ``Tor Vergata", I-00133 Roma, Italy
\item[25.]~Space Science Data Center - Agenzia Spaziale Italiana, Via del Politecnico, snc, I-00133, Roma, Italy
\item[26.]~Dipartimento di Fisica e Geologia, Universit\`a degli Studi di Perugia, I-06123 Perugia, Italy
\item[27.]~Istituto Nazionale di Fisica Nucleare, Sezione di Perugia, I-06123 Perugia, Italy
\item[28.]~INAF Istituto di Radioastronomia, I-40129 Bologna, Italy
\item[29.]~Grupo de Altas Energ\'ias, Universidad Complutense de Madrid, E-28040 Madrid, Spain
\item[30.]~Ruhr University Bochum, Faculty of Physics and Astronomy, Astronomical Institute (AIRUB), 44780 Bochum, Germany
\item[31.]~Department of Physical Sciences, Hiroshima University, Higashi-Hiroshima, Hiroshima 739-8526, Japan
\item[32.]~Friedrich-Alexander Universit\"at Erlangen-N\"urnberg, Erlangen Centre for Astroparticle Physics, Erwin-Rommel-Str. 1, 91058 Erlangen, Germany
\item[33.]~Dipartimento di Fisica, Universit\`a di Roma ``Tor Vergata", I-00133 Roma, Italy
\item[34.]~Max-Planck-Institut f\"ur Physik, D-80805 M\"unchen, Germany
\item[35.]~Universit\'e Paris Cit\'e, Universit\'e Paris-Saclay, CEA, CNRS, AIM, F-91191 Gif-sur-Yvette, France
\item[36.]~The George Washington University, Department of Physics, 725 21st St, NW, Washington, DC 20052, USA
\item[37.]~NASA Goddard Space Flight Center, Greenbelt, MD 20771, USA
\item[38.]~University of North Florida, Department of Physics, 1 UNF Drive, Jacksonville, FL 32224 , USA
\item[39.]~Yunnan Observatories, Chinese Academy of Sciences, 396 Yangfangwang, Guandu District, Kunming 650216, P. R. China
\item[40.]~Key Laboratory for the Structure and Evolution of Celestial Objects, Chinese Academy of Sciences, 396 Yangfangwang, Guandu District, Kunming 650216, P. R. China
\item[41.]~Universit\'e Bordeaux, CNRS, LP2I Bordeaux, UMR 5797, F-33170 Gradignan, France
\item[42.]~Department of Astronomy, University of Science and Technology of China, Hefei 230026, China
\item[43.]~School of Astronomy and Space Science, University of Science and Technology of China, Hefei 230026, China
\item[44.]~NASA Marshall Space Flight Center, Huntsville, AL 35812, USA
\item[45.]~Dipartimento di Fisica, Universit\`a di Trieste, I-34127 Trieste, Italy
\item[46.]~Istituto Nazionale di Fisica Nucleare, Sezione di Trieste, I-34127 Trieste, Italy
\item[47.]~Department of Astronomy, Department of Physics and Yale Center for Astronomy and Astrophysics, Yale University, New Haven, CT 06520-8120, USA
\item[48.]~Department of Physics and Center for Space Sciences and Technology, University of Maryland Baltimore County, Baltimore, MD 21250, USA
\item[49.]~Department of Physics, Faculty of Science, Mahidol University, Bangkok 10400, Thailand
\item[50.]~Hiroshima Astrophysical Science Center, Hiroshima University, Higashi-Hiroshima, Hiroshima 739-8526, Japan
\item[51.]~Vatican Observatory, Castel Gandolfo, V-00120, Vatican City State
\item[52.]~Department of physics and Astronomy, Louisiana State University, Baton Rouge, LA 70803, USA
\item[53.]~Istituto Nazionale di Fisica Nucleare, Sezione di Trieste, and Universit\`a di Trieste, I-34127 Trieste, Italy
\item[54.]~Department of Physics and Astronomy, University of Denver, Denver, CO 80208, USA
\item[55.]~INAF-Astronomical Observatory of Padova, Vicolo dell'Osservatorio 5, I-35122 Padova, Italy
\item[56.]~Institut f\"ur Astro- und Teilchenphysik, Leopold-Franzens-Universit\"at Innsbruck, A-6020 Innsbruck, Austria
\item[57.]~Departamento de F\'isica Te\'orica, Universidad Aut\'onoma de Madrid, 28049 Madrid, Spain
\item[58.]~Santa Cruz Institute for Particle Physics, Department of Physics and Department of Astronomy and Astrophysics, University of California at Santa Cruz, Santa Cruz, CA 95064, USA
\item[59.]~NYCB Real-Time Computing Inc., Lattingtown, NY 11560-1025, USA
\item[60.]~Purdue University Northwest, Hammond, IN 46323, USA
\item[61.]~Nagoya University, Institute for Space-Earth Environmental Research, Furo-cho, Chikusa-ku, Nagoya 464-8601, Japan
\item[62.]~Kobayashi-Maskawa Institute for the Origin of Particles and the Universe, Nagoya University, Furo-cho, Chikusa-ku, Nagoya, Japan
\item[63.]~Institute of Space Sciences (ICE, CSIC), Campus UAB, Carrer de Magrans s/n, E-08193 Barcelona, Spain; and Institut d'Estudis Espacials de Catalunya (IEEC), E-08034 Barcelona, Spain
\item[64.]~Instituci\'o Catalana de Recerca i Estudis Avan\c{c}ats (ICREA), E-08010 Barcelona, Spain
\end{enumerate}
$^{\dagger}$Corresponding Authors: Jian-Min Wang (wangjm@ihep.ac.cn), Jian Li (jianli@ustc.edu.cn)
}

\date{Feb 2023}
\begin{document}

\maketitle

\clearpage
\begin{abstract}
Relativistic jets around supermassive black holes (SMBHs) are well-known powerful $\gamma$-ray emitters.
In absence of the jets in radio-quiet active galactic nuclei (AGNs), how the SMBHs work in $\gamma$-ray bands is still unknown despite of great observational efforts made in the last 3 decades.
Considering the previous efforts, we carefully select an AGN sample composed of 37 nearby Seyfert galaxies with ultra-hard X-rays for the goals of $\gamma$-ray detections by excluding all potential contamination in this band.
Adopting a stacking technique, here we report the significant $\gamma$-ray detection (${\rm TS}=30.6$, or $5.2\,\sigma$) from the sample using 15-year Fermi-Large Area Telescope (LAT) observation.
We find an average $\gamma$-ray luminosity of the sample as $(1.5\pm1.0)\times10^{40}{\,\rm erg\,s^{-1}}$ at energies from 1-300\,GeV.
Limited by the well-known pair production from the interaction of $\gamma$-rays with low energy photons, $\gtrsim$ several GeV $\gamma$-rays are found to originate from an extended corona ($\sim 2.7\times 10^6\Rg$), whereas the canonical much more compact X-ray corona ($\sim 10\Rg$) is responsible for 1 to several GeV $\gamma$-rays.
The finding of the compact region lends to strong supports to the long-time theoretical expectations, but the extended corona is an unexpected finding.
One promising scenario is that the electron-positron pairs produced in the compact X-ray corona would expand as fireball, similar to that in $\gamma$-ray bursts, forming the structure of extended corona.
\end{abstract}

{Accretion disks of active galactic nuclei (AGNs) have hot coronae\cite{Bisnovatyi-Kogan1977,Galeev1979,Sunyaev1980,Bambic2024}, which are magnetically confined\cite{Inoue2018} ($\sim 10\,$Gauss) structures located a few $\Rg$ from the central supermassive black hole (SMBH)\cite{Fabian2009,Morgan2012,Reis2013}, where $\Rg=G\Mbh/c^2$ is the gravitational radius, $G$ is the gravitational constant, $\Mbh$ is the SMBH mass and $c$ is the speed of light.
They are composed of thermal electrons as evidenced by X-ray observations\cite{Haardt1993, Svensson1994, Witt1997} but also contain a population of non-thermal electrons\cite{Chael2017} whose existence has been revealed by radio observations\cite{Antonucci1988, Inoue2018}.
The non-thermal electrons in the coronae\cite{Wojaczyski2015,Inoue2019, Inoue2021,Gutierrez2021}, generated by shock acceleration\cite{Blandford1987,Inoue2019} and magnetic reconnection\cite{Sironi2014,Chael2017}, are predicted to produce $\gamma$-ray emission\cite{Inoue2019,Inoue2021,Romero2010}.}
Despite these expectations, several attempts to search for $\gamma$-ray emission from AGN coronae using EGRET\cite{Lin1993, Cillis2004} and {{\it Fermi}-LAT} data\cite{Teng2011, Ackermann2012_Seyfert} have led to non-detections, mainly due to the low $\gamma$-ray flux level and limited observation time.
In the present paper, we focused on radio-quiet  objects from the {\textit{Swift}}-BAT AGN Spectroscopic Survey (BASS) data release 2 sample\cite{Koss2022}.
This consideration arises from that these AGNs are generally ultra-hard X-ray bright (BAT bands: 14-195\,keV) and thus their coronae may potentially be $\gamma$-ray bright.
With 15-year accumulated {{\it Fermi}-LAT} data, we carried out a systematic search for $\gamma$-ray emission from the coronae of the radio-quiet BASS-selected AGNs.

\begin{figure*}
\centering
\includegraphics[scale=0.65, trim=10 30 0 0]{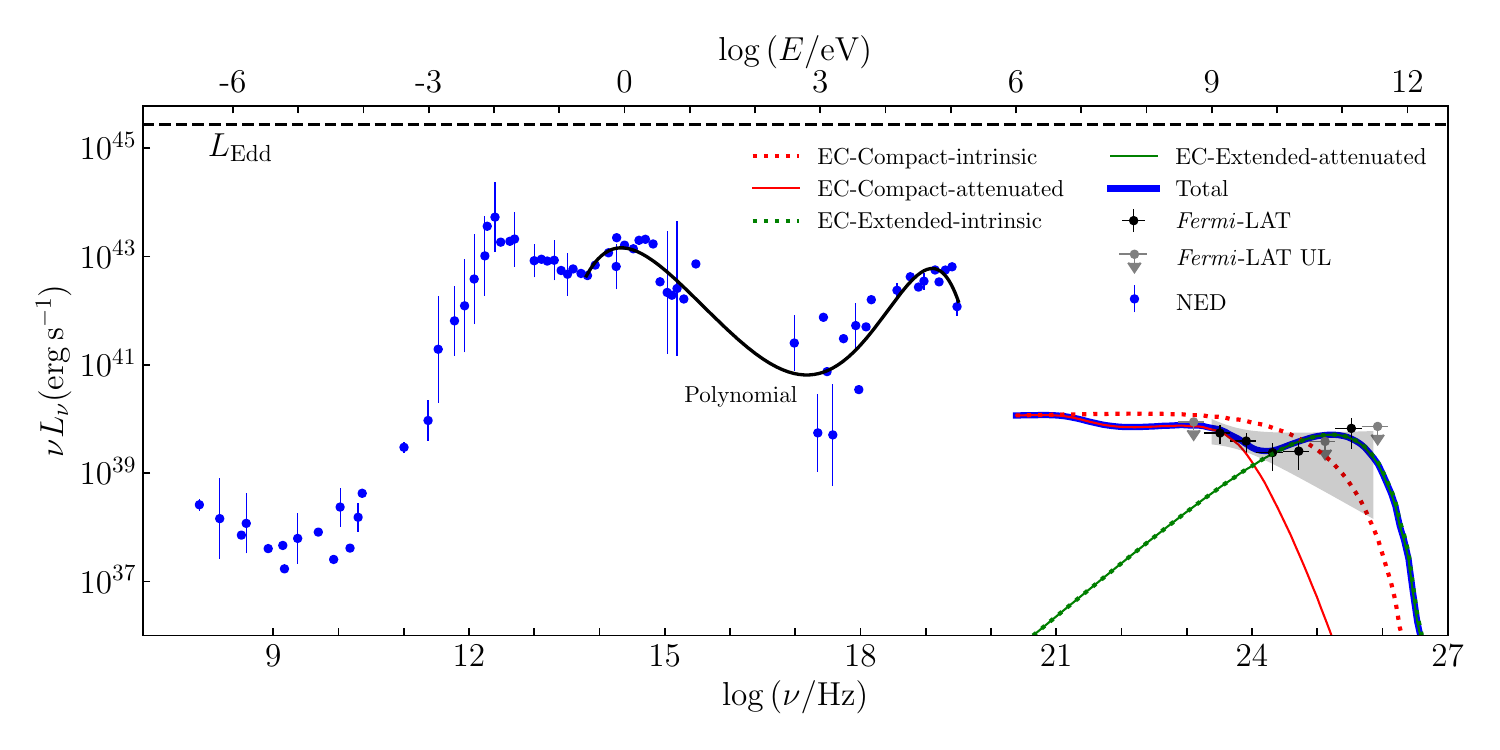}
\caption{\footnotesize
Average SED of the FGR sample constructed from archival data taken from the NASA/IPAC Extragalactic Database (NED) and our {\it Fermi}-LAT analysis.
The black and gray points are the $\gamma$-ray luminosities from this work (Supplementary Table \ref{table:TS_SED}).
The gray region represents the 1 $\sigma$ uncertainty of the $\gamma$-ray spectrum derived from the stacking result of $f_{\rm c} = 1.8^{+0.8}_{-0.8}\times 10^{-11}\, \fluxunit$ and $\Gamma = 2.32^{+0.40}_{-0.35}$.
All the points present average luminosities.
The SED from radio to X-rays for each source is normalized based on its hard X-ray luminosity and divided into 121 logarithmic bins with a width of 0.1 decade.
The black solid line represents the SED from the optical to the X-ray band used as seed photons for the inverse Compton scattering process,
which is smoothed using the polynomial curve fitting method.
Solid and dotted lines are the attenuated and unattenuated SED of the inverse Compton scattering for the compact (red) and extended corona (green),
while the solid blue line is the sum of the attenuated emission.
The black dashed line is the Eddington luminosity of  average SMBH mass $10^{7.3} \Msun$ on a logarithmic scale (see column 5 in Supplementary Table \ref{table:faint source}),
which are measured mainly from their broad Balmer lines or stellar velocity dispersions\cite{Koss2022}.
All error bars represent 1 $\sigma$ uncertainty.
}
\label{fig:SED_average}
\end{figure*}

We selected non-blazar, radio-quiet AGNs from the BASS sample, cross-matched them with the {\it Fermi}-LAT 4FGL-DR4 catalog, and analyzed their $\gamma$-ray data (see \S\,\ref{sect:sample} and \ref{sect:data analysis} in Methods).
This leads to a sample of 624 radio-quiet non-blazar AGNs which were below the {\it Fermi}-LAT sensitivity for individual $\gamma$-ray sources (see \S\,\ref{sect:sample} in Methods).
To search for weak $\gamma$-ray emission from these AGN, we analyzed a subsample adopting the stacking technique.
Since high X-ray fluxes represent high levels of AGN corona activity\cite{Haardt1993}, we aimed at nearby, ultra-hard X-ray bright emitters from the 624 radio-quiet non-blazar AGNs.
Sources located within 60\,Mpc with X-ray fluxes greater than $2\times 10^{-11} \,\efluxunit$ in the 14-195 keV {energy band} were further selected (see \S\,\ref{sect:sample} in Methods).
This subsample consisted of 37 nearby, ultra-hard X-ray bright, radio-quiet non-blazar AGNs,
which we refer to as the Faint $\gamma$-ray (FGR) sample (see \S\,\ref{sect:sample} in Methods).
We stacked their individual test statistic (TS) profiles in the 1-300\,GeV energy range to search for $\gamma$-ray emission and explore average properties.
As shown in Figure \ref{fig:TS_profile}, significant $\gamma$-ray emission is detected from the FGR sample with a maximum TS value of $30.6\,(5.2\,\sigma)$, corresponding to the best-fit result of $f_{\rm c} = 1.8^{+0.8}_{-0.8}\times 10^{-11}\, \fluxunit$ and $\Gamma = 2.32^{+0.40}_{-0.35}$,
which is consistent with the upper limit given in ref.\cite{Teng2011}.
The average distance of the FGR sample is 36.4\,Mpc with a standard deviation of 17.1\,Mpc, corresponding to an average luminosity of $\LGam =(1.5\pm 1.0)\times10^{40}{\,\rm erg\,s^{-1}}$.
The averaged multi-wavelength spectral energy distribution (SED) of the FGR sample is shown in Figure \ref{fig:SED_average}.

Several tests have been performed to rule out origins other than the AGN coronae of the $\gamma$-ray emission observed in the FGR sample.
We constructed a control sample consisting of $27$ sources selected from the BASS sample using the same criteria as those used for the FGR sample but with an integrated X-ray flux below $2\times 10^{-11} \,\efluxunit$, which represents a comparably low level of AGN corona activity.
No significant $\gamma$-ray emission was detected (TS = $2.8$, or $1.2\, \sigma$,
see \S\,\ref{sect:result} in Methods), 
yielding a flux upper limit of $1.86\times 10^{-11}\fluxunit$ at 95\% confidence level in the 1-300\,GeV energy band.
Additionally, 37 empty sky positions were randomly selected to perform a background stacking analysis, which leads to no $\gamma$-ray detection,  demonstrating that the detected $\gamma$-ray emission from FGR sample does not arise from background fluctuations (see \S\,\ref{sect:result} in Methods).
Apart from the corona, $\gamma$-rays from radio-quiet AGN may originate from star formation\cite{Ackermann2012_SF, Ajello2020}, low-power jets, or AGN-driven outflows\cite{Ajello2021, McDaniel2023}.
We have carried out tests and demonstrated that the $\gamma$-ray contributions from these processes are most likely negligible for the FGR sample (see {\S\,\ref{sect:other_origin}} in Methods).

\begin{figure*}
\centering
\includegraphics[scale=0.99, trim=0 30 0 0, clip]{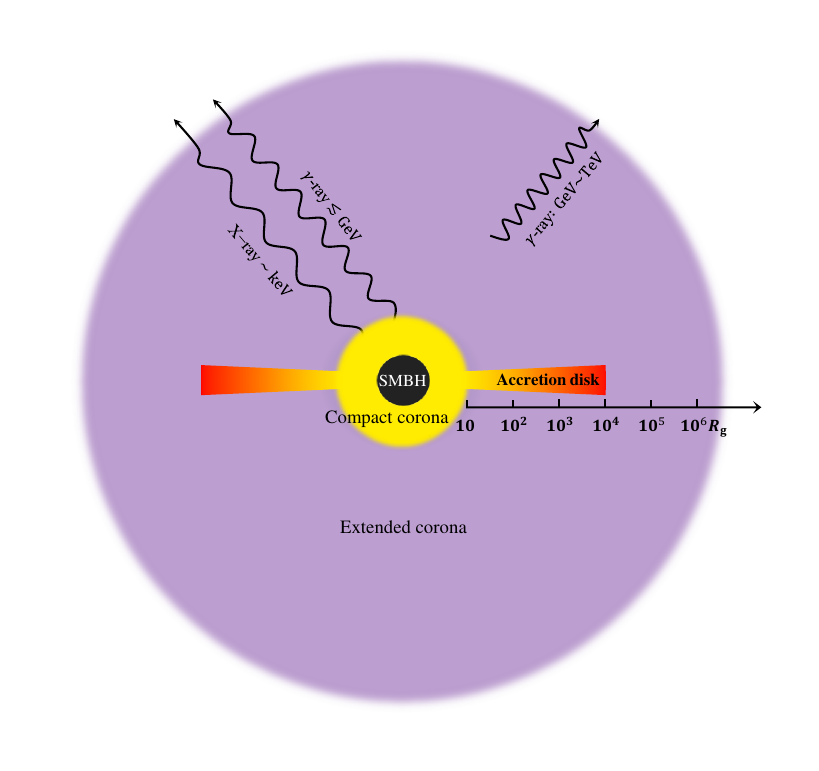}
\caption{\footnotesize
Illustration of the AGN corona scenario used to explain the $\gamma$-ray emission from the FGR sample. The extended corona ($\gtrsim 2.7\times 10^6\,\Rg$ or $2.8\,\rm pc$) is responsible for $\gtrsim\,$ several GeV emission whereas the compact corona ($10\,\Rg$) for $\lesssim\,$ several GeV photons.
}
\label{fig:cartoon}
\end{figure*}

The corona geometry is still a matter of debate with different scenarios {being proposed}, including the lamp-post corona\cite{Martocchia1996, Miniutti2004}, plane-parallel and hemispherical model\cite{Petrucci2000}, and patchy corona\cite{Wilkins2015}.
A spherical geometry, as the simplest model, is adopted in this paper {(Figure \ref{fig:cartoon})}.
In the corona, electrons can be accelerated to relativistic velocities by shocks\cite{Blandford1987, Inoue2019} or magnetic reconnection\cite{Sironi2014, Chael2017}.
Intrinsic $\gamma$-ray emission would be expected from the inverse Compton scattering process between the non-thermal electrons and optical, ultraviolet (UV) photons (see \S\,\ref{sect:discussion} in Methods).
On the other hand, the $\gamma$-rays produced in the corona would be attenuated due to pair production ($\gamma\,\gamma^{\prime}\rightarrow e^{+} e^{-}$) interactions.
For the FGR sample, the average dimensionless soft X-ray luminosity ($\lamX =\Lsoftx/\Ledd \sim 4.8\times 10^{-4}$, see \S\,\ref{sect:discussion} in Methods) is comparably high,
where $\Ledd=4\pi G \Mbh m_{\rm p} c /\sigmaT$ is the Eddington luminosity,
$m_{\rm p}$ is the proton rest mass, and $\sigmaT$ is the Thomson scattering cross section. 
Thus, $\gamma$-ray photons $\gtrsim$ several GeV would be largely absorbed within AGN corona (a few $\Rg$)\cite{Fabian2015, Kamraj2022} and rarely escape.
It is inconsistent with the observed $\gamma$-ray spectrum of FGR sample extending beyond 10\,GeV (Figure \ref{fig:SED_average}).
In order to understand the observed gamma-rays from FGR sample with pair production absorption incorporated, we have to consider two regions of the hot corona for the current situation (Figure \ref{fig:cartoon}).
In the compact region of corona ($\sim10\,\Rg$, “compact corona” hereafter), thermal electrons dominate and produce a high density of X-ray photons,
which leads to severe absorption of $\gamma$-ray photons $\gtrsim$ several GeV via pair production (see solid and dotted red lines in Figure \ref{fig:SED_average}).
Non-thermal electrons in the compact corona regions mainly contribute $\gamma$-ray photons at $\lesssim$ several GeV (Figure \ref{fig:SED_average}), which is consistent with theoretical models\cite{Inoue2019}.

To explain the $\gamma$-ray spectrum $\gtrsim$ several GeV of the FGR sample (Figure \ref{fig:SED_average}) as most likely corona origin, we propose that the non-thermal electrons responsible for $\gamma$-ray emission occupy a much larger region than the compact corona, forming the “extended corona” (Figure \ref{fig:cartoon}), which is beyond current theoretical models\cite{Inoue2019}.
In the extended corona, the non-thermal electrons dominate,
and the X-ray photon density decays proportionally to $R^{-2}$, where $R$ is the corona size. 
The pair production is alleviated in the extended corona.
The observed $\gamma$-ray photons $\gtrsim$ several GeV mainly originate from this region (Figure \ref{fig:SED_average}).
To account for the high-energy $\gamma$-ray emission, the extended corona is estimated to be at least $\sim2.7\times 10^6\,\Rg$ (see Table \ref{table:sed_para}),
which is much greater than the compact corona (see \S\,\ref{sect:discussion} in Methods). 
The energy of the non-thermal electrons in the compact corona is $\xicom=1.8\times 10^{-2}$ that of the thermal electrons. This result is consistent with the limiting values found in ref.\cite{Ackermann2012_Seyfert}.
The number density of non-thermal electrons in the extended corona ($\Next={6.5\times 10^{-7}}\,\rm cm^{-3}$) is much lower than that in the compact corona ($\Ncom={7.6}\times10^7\,\rm cm^{-3}$).

While the observations may hint at the presence of two distinct regions of the X/$\gamma$-ray corona, how such a structure would form remains unclear at present.
Actually, there is evidence for the expansion of coronae in AGN\cite{Kara2023}, implying that they do not hold static equilibrium.
Expanding coronae are very different from static coronae\cite{Haardt1993, Svensson1994, Witt1997}.
Expansion in coronae could be driven by pair production which would expand as fireball, similar to that in $\gamma$-ray bursts. This bridges the extended and compact coronae (see an extensive review on the physics of pair production\cite{Svensson1986}).
In such a scenario, electrons can be re-accelerated by shocks from the ambient medium during the expansion.
Clearly, the coronae expansion process should be explored in detail to compare the theoretical model of pair plasma with observed $\gtrsim$ several GeV spectra.
Future observations of the FGR sample, from hundreds of GeV to TeV may enable further constraints {to be placed on possible scenarios} and provide information on the SMBH (e.g., spin\cite{Wang2008}).

In summary, by adopting a stacking technique, we have detected a $5.2\,\sigma$ signal consistent with the long expected $\gamma$-ray emission from AGN coronae. We proposed that the non-thermal electrons responsible for the $\gamma$-ray emission are distributed over a much larger region ($\sim 2.7\times 10^6\,\Rg$) as an extended part of a hot corona compared with the compact corona ($\sim10\,\Rg$) in the AGN. Co-existence of the compact and the extended parts of the corona supports a scenario where the hot corona has an expanding configuration.
We also note that $\gamma$-ray emissions and neutrinos contributed by energetic proton-proton collisions in hot corona could be unignorable\cite{Inoue2019}, or produced by nuclear star clusters with typical sizes ($\lesssim 10\,$pc)\cite{Neumayer2020} of AGN host galaxies.
Future variability studies of the $\gamma$-ray and radio emissions from the coronae of AGN will advance the understanding of their evolution in time as well as the role of pair production in such systems.
Further high-energy measurements of AGN coronae in the TeV regime (e.g. by the Large High Altitude Air Shower Observatory (LHAASO)\cite{2022ChPhC} and the Cherenkov Telescope Array (CTA)\cite{Actis2011}) will extend the spectrum and put better constraints on the maximum energy of accelerated particles.

\clearpage

\part*{Methods}

\section{Sample Selection} 
\label{sect:sample}

We start from the BASS DR2 sample, which contains 858 hard X-ray selected sources (Supplementary Figure \ref{fig:flowchart}).
We first remove all the blazars labeled with BZQ (beamed AGNs with broad lines),
BZG (beamed AGNs hosting galaxy but lacking broad lines), or BZB (traditional continuum-dominated blazars with no emission lines or host galaxy features), which are 105 sources.
We then use two blazar catalogs, the Roma-BZCAT catalog\cite{Massaro2015} and the WISE Blazar-like Radio-Loud Sources (WIBRaLS) catalog\cite{Abrusco2019}, to search for blazars that are spatially coincident (to within $5^{\prime\prime}$) with sources in BASS sample.
The Roma-BZCAT catalog consists of 3561 sources confirmed as blazars based on their properties in the radio, optical, and X-ray bands.
The WIBRaLS catalog includes 9541 blazar-like sources observed by the Wide-Field Infrared Survey Explorer (WISE).
WISE maps the whole sky in mid-infrared bands centered at wavelengths of 3.4, 4.6, 12, and 22 $\mu$m (see ref.\cite{Wright2010}).
Using the above two blazar {catalogs}, we exclude sources that are associated or spatially coincident with confirmed (11 sources) and blazar-like sources (10 sources).
After this first selection, there remains 732 sources.
The 10 blazar-like sources all have distance larger than 60 Mpc and would not be included in the final FGR sample if considered in following sample selection steps.
%

We exclude sources that are associated with non-blazar radio-loud AGN
since any detected $\gamma$-ray emission may come from the jet
that produces the radio emission.
To remove radio-loud AGN, we eliminate radio-loud sources that are spatially coincident (to within $5^{\prime\prime}$) with sources from radio {catalogs}.
For this work, we mainly use the Very Large Array Sky Survey (VLASS)\cite{Gordon2021} as it is the newest radio sky survey in the $S$ band (2{-}4 GHz).
This survey began in 2017 and covers the entire northern sky at declinations above $-40^\circ$ with an angular resolution of $2.5^{\prime\prime}$.
Additional radio {catalogs} used in this work include the ongoing LOw-Frequency ARray (LOFAR) Two-metre Sky Survey (LoTSS)\cite{Shimwell2022},
the Faint Images of the Radio Sky at Twenty-centimeters (FIRST)\cite{Helfand2015},
the Sydney University Molonglo Sky Survey (SUMSS)\cite{Mauch2003},
the NRAO VLA Sky Survey (NVSS)\cite{Condon1998},
the Westerbork Northern Sky Survey (WENSS)\cite{Rengelink1997},
and the Parkes Radio Sources catalog (PKSCAT90)\cite{Wright1990}.
A typical parameter used to classify AGNs as radio loud is the classical radio-loudness parameter, $\Rrb \equiv \LnuV /L_{\rm B}$,
where $\LnuV$ is the radio luminosity at 5 GHz and $L_{\rm B}$ is the optical luminosity\cite{Elvis1994, Sikora2007}.
However, for Seyfert galaxies, the value of $\Rrb$ depends strongly on the subtraction of host galaxy emission\cite{Ho2001}.
Following ref.\cite{Ackermann2012_Seyfert}, we define the so-called ``hard X-ray radio-loudness parameter” given by
\begin{equation}
\label{eq:radio loudness}
\Rrx \equiv \frac{[\nu L_{\nu}]_{\nu=\rm 3\,GHz}}{\Lx},
\end{equation}
where $L_{\nu}$ is monochromatic radio luminosity and 
$\Lx$ is the hard X-ray luminosity integrated from 14 to 195 keV.
We use the 3 GHz radio luminosity since the mean frequency of the VLASS survey is 3 GHz (see ref.\cite{Lacy2020}).
We transform radio luminosities at other bands to 3 GHz luminosities using $L_\nu \propto \nu^{-0.71}$ (see ref.\cite{Gordon2021}).
The critical $\Rrx$ value is set as $10^{-4}$,
which roughly corresponds to classical radio-loudness of $\Rrb=10$ (see refs.\cite{Terashima2003, Panessa2007, Ackermann2012_Seyfert}).
As can be seen in Supplementary Figure \ref{fig:histRrx}, $\Rrx$ is a well defined parameter that can exclude most radio-loud sources.
To further exclude radio-loud sources and get a purer sample, the classical radio loudness $\Rrb<10$ is also adopted in sample selection\cite{Miller1990,Visnovsky1992,Kellermann1994}.
Using $L_\nu \propto \nu^{-0.71}$ (see ref.\cite{Gordon2021}), we transform radio luminosities from 3 GHz to 5 GHz.
The $B$ band luminosity is derived from the ref.\cite{Veron2010}.
After this second selection, there remains 634 sources.
We would like to note that NGC 4945, which is a 4FGL source and recently detected with significant low-energy $\gamma$-ray emission down to 20 MeV\cite{Murase2024}, is not included in these 634 sources.
Its hard X-ray radio-loudness parameter is larger than the critical value and $\gamma$-rays from NGC 4945 may be attributed to star-forming activities (see refs.\cite{Wang2004a,Lenain2010,Teng2011,Murase2024}).

To identify $\gamma$-ray sources, we spatially crossmatch the current 634 sources from the BASS sample with sources from the {{\it Fermi}-LAT} 4FGL-DR4 catalog (which contains 7194 sources)\cite{Ballet2023},
determining the crossmatching region radii from the localization error of $\gamma$-ray position ($\tt Conf\_95\_SemiMajor$ parameter).
7 sources are found with significant $\gamma$-ray emission (TS $>25$).
As a complement, another 2 sources with close $\gamma$-ray {neighbors} (i.e., within a separation of $0.08^\circ$; see ref.\cite{Tsuji2021}) are also counted. 
9 sources in total are listed in Supplementary Table \ref{table:jet contamination}.
A detailed discussion on the results of crossmatching the BAT 105-month catalog\cite{Oh2018} with the 4FGL-DR2 catalog has been presented in ref.\cite{Tsuji2021}.
However, we note that the $\gamma$-ray emission associated with these 9 sources may not necessarily be connected with their coronae
(see discussion in \S\,\ref{sect:cross match}).
After this cross match procedure, 625 sources remains.

We then perform a {\it Fermi}-LAT analysis
(see \S\,\ref{sect:data analysis})
for these 625 sources individually to search for any $\gamma$-ray emission.
Possible $\gamma$-ray emission is seen from NGC 3281 but could be the result of a background fluctuation
(see \S\,\ref{sect:NGC result}).
After this procedure, 624 sources remain.

To search for weak $\gamma$-ray emission from the remaining sources, we use the stacking technique (see \S\ref{appdx: stacking}) to analyze a subsample of nearby radio-quiet AGNs.
In constructing this sub-sample, we select nearby, {ultra-hard X-ray bright} emitters.
We first restrict the sample to those sources that are located within $D_{\rm max}=60$\,Mpc.
This critical distances are derived by the {\it Fermi} threshold ($F_{\rm threshold}\sim 10^{-13}\,{\rm erg\,cm^{-2}\,s^{-1}}$) (https://www.slac.stanford.edu/exp/glast/groups/canda/lat\_Performance.htm) and theoretical expectations of $\gamma$-ray emissions ($L_{\gamma}\sim 4\times 10^{40}\,\rm erg\,s^{-1}$)\cite{Inoue2021}.
We derived $D_{\rm max}=\left(L_{\gamma}/4\pi F_{\rm threshold}\right)^{1/2}\approx 60\,$Mpc.
72 sources remain after the distance cut.

We also exclude star forming galaxies (SFGs) to remove {any $\gamma$}-ray contribution from star formation activity\cite{Ackermann2012_SF, Ajello2020}.
We compare the 72 sources with the SFG catalog in ref.\cite{Ackermann2012_SF} to locate SFGs.
The Baldwin-Phillips-Terlevich (BPT) diagram\cite{Baldwin1981,Oh2022} with the emission-line ratios of $\log \rm \,([OIII]\lambda 5007/H\beta)$ and $\log \rm \,([NII]\lambda 6583/H\alpha)$ is also used to locate SFGs (http://www.bass-survey.com/dr2.html).
%
We exclude 6 possible SFGs, NGC 1365, UGC 6728, ESO 424-12, NGC 7479, CTS 103, and Fairall 346, and thus 66 sources remained.
For each of these (66) sources, we carried out a counterpart check with X-ray and other wavelengths to search for possible source confusion.
The X-ray source SWIFT J0209.5-1010 has two possible infrared counterparts, NGC 833 and NGC 835.
To avoid source confusion, NGC 833 and NGC 835 are excluded from the subsample.
After this exclusion, 64 sources remain.
High X-ray fluxes represent high level of AGN corona activities.
The source distribution against X-ray flux in the 14-195 keV of these 64 sources is shown in Supplementary Figure \ref{fig:source}{\textbf{a}}.
We limit the X-ray flux in the 14-195 keV energy band greater than $2\times 10^{-11} \,\efluxunit$, which is at the peak of the source distribution.
It is slightly lower than the criterion used in ref.\cite{Ackermann2012_Seyfert} but exceeds 2 times the {\it Swift}-BAT threshold\cite{Oh2018}.
After the hard X-ray flux cut, 37 sources remain.
In \S\,\ref{sect:result}, we discuss the dependence of the results on the hard X-ray flux cut.

After following the above procedure, our sub-sample consists of 37 nearby, non-blazar, {non-SFG}, {ultra-hard} X-ray bright, radio-quiet AGNs.
We call this sub-sample the Faint $\gamma$-ray (FGR) sample (Supplementary Table \ref{table:faint source}).
We also tested a more stringent cut on $\Rrx$ as $5\times 10^{-5}$ (or $10^{-5}$).
The FGR sample would contain 36 (or 27) sources, leading to a stacked TS of $31.3$ (or 20.9).

\section{{\it Fermi}-LAT data analysis}
\label{sect:data analysis}

\subsection{Likelihood analysis}

LAT on the Fermi Gamma-ray Space Telescope scans the whole sky\cite{Atwood2009}.
High-energy photons ranging from 20 MeV to above 300\,GeV are captured and undergo pair-conversion when going through the converter foils. 
The detection of photons is characterized by the Poisson process, therefore, {the binned} likelihood method (https://fermi.gsfc.nasa.gov/ssc/data/analysis/scitools/) is used to analyze the LAT data.
The probability of {observing} $m_i$ photons {in the $i$ th bin} can be described by a Poisson distribution (https://fermi.gsfc.nasa.gov/ssc/data/analysis/documentation/Cicerone/
Cicerone\_Likelihood/):
\begin{equation} \label{eq:probability}
P_{\rm i}=\frac{n_i^{m_i}}{m_i!} \, {\rm exp}\left(-n_i\right),
\end{equation}
where $n_i$ is the expected number of photons decided by the model prediction, for example, a power-law (see Equation \ref{eq:powerlaw}) is used in this work. 
The likelihood can be written as:
\begin{equation} \label{eq:likelihood}
\mathcal{L}={\rm exp}\, (-N) \prod_i \frac{n_i^{m_i}}{m_i!},
\end{equation}
where $N=\sum n_i$ is the total expected photon number {from} all the bins. 
As mentioned above, $\mathcal{L}$ depends on both models ($n_i$) and observations ($m_i$). 
The best model parameters are expected to give a maximum of $\mathcal{L}$.

Furthermore, TS is used to present the significance of any source detection,
which is defined as
\begin{equation}
\rm TS=2\,( log \mathcal{L}_{1,\rm max}-log\mathcal{L}_{0,\rm max}),
\end{equation}
where $\mathcal{L}_{0,\rm max}$ and $\mathcal{L}_{1,\rm max}$ are the maximum likelihood {values} in the null and alternative hypothesis, respectively\cite{Mattox1996}. 
The null hypothesis means that there is no source at the given position while the alternative hypothesis means that the model with the source is preferred.
According to Wilks Theorem, TS is asymptotically distributed as $\chi ^{2}(\rm d.o.f.)$ in the null hypothesis\cite{Wilks1938,Mattox1996},
where the degree of freedom ($\rm d.o.f.$) is equal to the number of model parameters for additional source, here $\rm d.o.f.=2$ for the power-law model.

\subsection{Stacking technique} \label{appdx: stacking}

The stacking technique is often used to explore $\gamma$-ray average properties for astrophysical populations. 
58 nearby X-ray-bright galaxy clusters images were stacked to search for $\gamma$-ray emission, resulting in a $2\,\sigma$ flux upper limit\cite{Reimer2003}. 
Additionally, the stacking analysis of 112 extended clusters revealed a bright $\gamma$-ray ring at the viral shock position\cite{Reiss2018}. 
For individual weak $\gamma$-ray sources, the TS is generally too low ($<25$) to give a significant detection. 
While for weak $\gamma$-ray source populations, it is useful to perform a stacking analysis,
which enhances the signal-to-background ratio.

Several stacking algorithms have been developed to search {\it Fermi}-LAT data for $\gamma$-ray emission\cite{Huber2012,Ackermann2011,Paliya2019,Ajello2021}. 
In this work, we adopt the stacking method introduced based on the individual source likelihood profiles and the additivity of log-likelihood\cite{Paliya2019}.
In this case, $i$ in Equation (\ref{eq:likelihood}) traverses all the bins and sources,
which means that $\mathcal{L}$ of different sources can be multiplied and the values of the TS can be added together.
For each source in a population, TS values are calculated in the bins of flux $\&$ spectral index parameter space, forming a TS profile.
TS profiles of individual source in a population are stacked and added, leading to a summed TS profile, from which the average spectral properties of the population can be estimated.
It is convenient to add new sources to populations since all the TS profiles are generated independently with likelihood analysis, which allows for combinations of any set of sources.
Furthermore, this method was improved by dividing the photons into four types, corresponding to four point-spread functions (PSFs)\cite{Ajello2021} of {\it Fermi}-LAT, which can lead to a more accurate maximum log-likelihood. 
As considered above, the likelihood profile stacking technique in ref.\cite{Ajello2021} is adopted in this work. 

\subsection{Systematic $\gamma$-ray search}

In the systematic $\gamma$-ray search process, we check if any of the 625 targets (Supplementary Figure \ref{fig:flowchart}) were individually detected using 15 years of {\it Fermi}-LAT data from modified Julian date (MJD) 54683 to MJD 60371 (August 4, 2008 to March 2, 2024).
We extract and analyze the data using the ScienceTools (v2.0.8) and Fermipy (v1.1.6) package\cite{Wood2017}
(https://fermipy.readthedocs.io/en/latest/).
In the first step, we select source class events with the $\tt{gtselect}$ tool by setting $\tt evclass=128$ and $\tt evtype= 3$.
A $15^{\circ}$ regions of interest (ROIs) of data around each individual source is adopted (https://fermi.gsfc.nasa.gov/ssc/data/analysis/).
We use a zenith angle of $90^\circ$ to reduce contamination from the Earth's limb.
We include all sources from the 4FGL-DR4 catalog ($\tt gll\_psc\_v32$)\cite{Ballet2023} that are located within $20^{\circ}$ of our sources.
Second, we use the $\tt gtmktime$ filter to select good time intervals (GTI) and valid data by setting $\tt(DATA\_QUAL>0)\&\&(LAT\_CONFIG==1)$.
The photons are divided into 30 logarithmic energy bands between $\emin$ and $\emax$.
The galactic diffuse emission ($\tt gll\_iem\_v07$) and isotropic emission ($\tt iso\_P8R3\_SOURCE\_V3\_\tt v1$) models provided by the {{\it Fermi}-LAT} team are used in the analysis.

We model the $\gamma$-ray spectrum {of each source in the FGR sample} using a power-law model
\begin{equation}\label{eq:powerlaw}
\frac{dN}{dE} =N_0\left(\frac{E}{E_0}\right)^{-\Gamma},
\end{equation}
 where $N_0$ is the normalization of the $\gamma$-ray flux,
$E_0$ is the fiducial energy,
and $\Gamma$ is the spectral index.
In this analysis, we fix $E_0$ to {2000} MeV and leave $N_0$ and $\Gamma$ free to vary.
We leave free to vary the model parameters for sources within $5^{\circ}$ of the target source as well as those for the diffuse backgrounds.
We implement the energy dispersion correction for every energy bin by setting $\tt edisp\_bins=-1$.
Finally, we perform the likelihood fit using the fit method with the $\tt MINUIT$ optimizer and setting the tolerance to $10^{-4}$ (see ref.\cite{James1975}).
TS maps are produced using the $\tt tsmap$ algorithm in Fermipy {and used to check the background residuals}.

\subsection{Stacking analysis of the FGR sample}

In the analysis for the 37 sources in the FGR sample, there are three differences from the above analysis, {adopted for faint source stacking analysis.}
Firstly, the photon energy range is set to $\emid$ and $\emax$ 
in order to minimize source confusion due to background photons, since LAT has a smaller PSF above 1\,GeV.
Secondly, following the method presented in refs.\cite{Paliya2019, Ajello2021},
we perform a joint likelihood analysis of
photon data with varying degrees of quality for the reconstructed direction.
The photon data {are} divided into four quality quartiles from the lowest-quality quartile (PSF0) to the best-quality quartile (PSF3).
We set ${\tt evtype}= i$ where $i$ = 4, 8, 16, or 32 and use their corresponding isotropic emission model $\tt iso\_P8R3\_SOURCE\_V3\_PSF\it k\_\tt v1$ where the value of $k$ (0, 1, 2, or 3) corresponds to one of the four PSFs.

Thirdly, we consider possible contamination from nearby blazars.
Although we have excluded blazars in our sample (see \S\,\ref{sect:sample}),
considering the PSF of LAT ($\sim 0.85^{\circ}$ at 1\,GeV) (https://www.
slac.stanford.edu/exp/glast/groups/canda/lat\_Performance.htm),
other blazars in the vicinity of our source position may introduce contamination.
To evaluate the contamination effect, we match the FGR sample against Roma-BZCAT\cite{Massaro2015},
adopting a radius of $\sim 0.85^{\circ}$.
There are 16 blazars near 11 FGR sources as shown in Supplementary Table \ref{table:nearbyBlazar}.
The TS values of the 11 FGR sources range from 11 to 0 (Supplementary Table \ref{table:nearbyBlazar}).
Among the 16 blazars, 8 are spatially associated with 4FGL-DR4 sources and thus have already been included in the background model.
Dedicated {\it Fermi}-LAT analysis of the remaining 8 blazars leads to no detection.
However, to exclude their potential contamination to the FGR sources,
new point sources at blazar optical positions are added to the background model in the data reduction process.
$\gamma$-ray photons from these blazars, if any, would be fitted $\&$ attributed to these blazars and would not contaminate the FGR sample.

We use a power-law spectral model (see Equation \ref{eq:powerlaw}) to {characterize} the $\gamma$-ray spectra of the individual sources in the FGR sample.
The counts flux $f_{\rm c}$ can be calculated by integrating the model from $E_{\rm min}$ to $E_{\rm max}$.
Since $N_0$ and $E_0$ in Equation (\ref{eq:powerlaw}) are degenerate,
there are only two degrees of freedom, $f_{\rm c}$ and $\Gamma$.
$f_{\rm c}$ is divided into 40 logarithmic bins from $10^{-14}$ to $10^{-10} \,\fluxunit$,
and $\Gamma$ is divided into 28 linear bins from 1.1 to 3.9. 
The lower flux limit in this analysis is about three orders of magnitude below the {{\it Fermi}-LAT} sensitivity for individual $\gamma$-ray sources at energies $>1$\,GeV ($\sim 10^{-11} \,\fluxunit$)\cite{Atwood2009}.
We chose this lower limit in order to have a value that is small enough to represent the absence of a $\gamma$-ray source\cite{Paliya2019},
which is the null hypothesis used to calculate the TS.

In constructing the background model, we {search for} new $\gamma$-ray sources in the ROI with the $\tt find\_sources$ algorithm.
We include only sources with TS $>$25 (setting the $\tt {sqrt\_ts\_threshold}$ parameter to 5),
and we set the minimum separation between sources $\tt {min\_separation}$ to $0.5^{\circ}$.
We then use the $\tt fit$ algorithm to optimize the model parameters including indexes $\Gamma$ and fluxes $f_{\rm c}$ of sources within $5^{\circ}$.
The derived TS and upper limit of luminosity for each FGR source are listed in Supplementary Table \ref{table:faint source}.
None of the 37 sources are individually {detected} ($\rm TS<25$);
therefore, we stack the sources with the likelihood profile stacking technique in refs.\cite{Paliya2019, Ajello2021} in order to enhance their signal-to-background ratio,
allowing us to determine whether they collectively exhibit significant $\gamma$-ray emission and to infer their average $\gamma$-ray properties.

We then perform a likelihood profile analysis with the $\tt profile$ algorithm by setting an array of normalizations derived from the 40 logarithmic bins of $f_{\rm c}$ and each corresponding index $\Gamma$. 
Note that all of the model parameters of the background sources are now fixed to those found in the previous analysis except for diffuse sources (GALACTIC, ISOTROPIC, and any other extended sources in the ROI) to speed up the optimization when getting the TS profile\cite{Paliya2019, Ajello2021}. 
We obtain TS profiles for each individual source, and then stack them to generate the total TS profile for the FGR sample (Figure \ref{fig:TS_profile}).
We use the $\tt tsmap$ algorithm in the Fermipy package to generate {a} TS map with $0.1^{\circ}$ pixel size for the individual sources and then the background residuals are checked.

\section{Results of the stacking analysis of FGR sample}
\label{sect:result}
\subsection{Significant detection.}

We plot the TS distribution from the $>$ 1\,GeV analysis for the 37 sources in the FGR sample in Supplementary Figure \ref{fig:TS_distribution}.
The TS values for these sources are provided in Supplementary Table \ref{table:faint source}.
In our analysis, with $\rm d.o.f.=2$ for the power-law model, the TS distribution of the FGR sample would follow a $\chi^2(\rm d.o.f.=2)$ distribution (dotted line in Supplementary Figure \ref{fig:TS_distribution}) if they were purely background fluctuations.
However, from TS$>$3 the FGR sample distribution exceeds the number of source expected from $\chi^2$(d.o.f.=2) distribution (Supplementary Figure \ref{fig:TS_distribution}).
The expected number of source with TS larger than 3 is 8.3 based on the $\chi^2(\rm d.o.f.=2)$ distribution for a sample containing 37 sources.
But in the FGR sample, there are 16 sources with TS larger than 3,
which demonstrates that the TS values of our FGR sample cannot be explained as  background fluctuations.

\begin{figure*}
\centering
\includegraphics[scale=0.9,trim= 30 10 0 0]{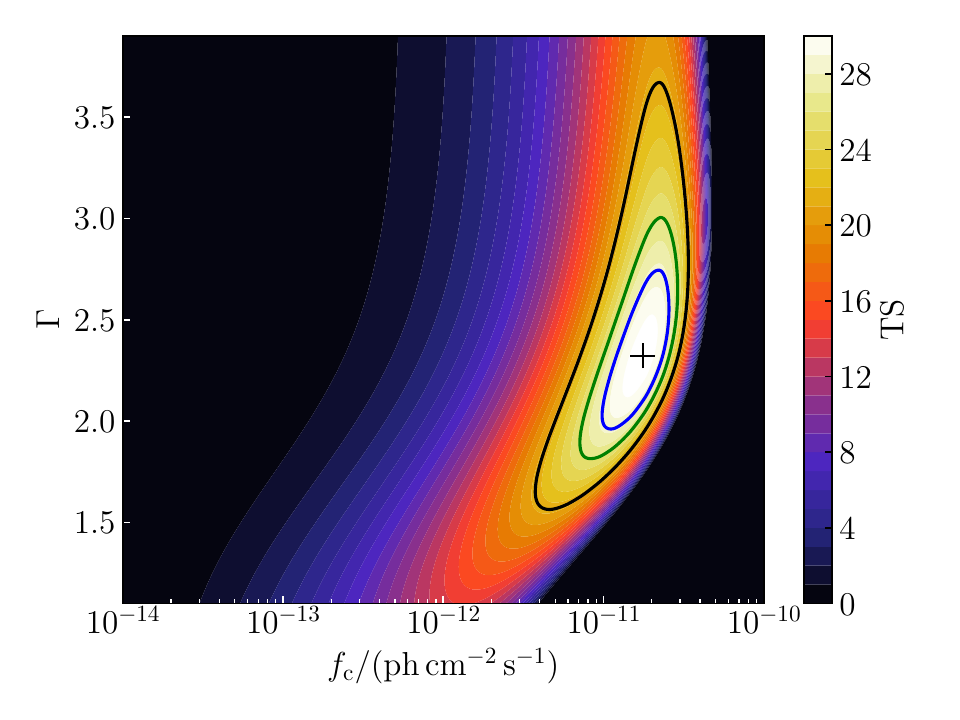}
\caption{\footnotesize
Stacked TS profile for FGR sample containing 37 sources. 
The TS value is color-coded for each flux and index combination.
The maximum TS value is $30.6\,(5.2\,\sigma)$, corresponding to the best-fit result of $f_{\rm c} = 1.8^{+0.8}_{-0.8}\times 10^{-11}\, \fluxunit$ and $\Gamma = 2.32^{+0.40}_{-0.35}${, marked by the black cross}.
The three solid contours represent the 68\%, 90\% and 99\% confidence level.}
\label{fig:TS_profile}
\end{figure*}

Therefore, we stack the profiles together to explore their average properties, as shown in Figure \ref{fig:TS_profile}.
The maximum TS value is $30.6\,(5.2\,\sigma)$, corresponding to the best-fit result of $f_{\rm c} = 1.8^{+0.8}_{-0.8}\times 10^{-11}\, \fluxunit$ and $\Gamma = 2.32^{+0.40}_{-0.35}$.
This result is consistent with the upper limit given in ref.\cite{Teng2011}.
The average distance is $36.4\,(\pm 17.1)$ Mpc, where the uncertainty is the standard deviation of the distribution, corresponding to an average luminosity of $\LGam =(1.5\pm 1.0)\times10^{40}{\,\rm erg\,s^{-1}}$.
We also calculate the TS-weighted isotropic $\gamma$-ray luminosity given by
\begin{equation}
\label{eq:lum_TSwighted}
L_{\gamma}^{\rm TS}=\frac{\sum_{i=1}^{37}\,L_{\gamma,i}\times {\rm TS}_{i}}{\rm TS_{tot}},
\end{equation}
where $i$ indicate the $i$th source,
$L_{\gamma,i} = 4\pi D_{i}^{2}f_{\rm e}$,
$D_{i}$ is the distance,
{${\rm TS}_{i}$ is the fitting result (Supplementary Table \ref{table:faint source}),}
$f_{\rm e}$ is the energy flux of the stacking result,
and $\rm TS_{tot}$ is the summed TS value.
For the FGR sample, we calculate a value of $\LGam^{\rm TS}=(1.9\pm 1.3)\times10^{40}{\,\rm erg\,s^{-1}}$ for the TS-weighted isotropic $\gamma$-ray luminosity,
which is comparable to the non-weighted average luminosity.
To derive the $\gamma$-ray SED, We repeat the stacking analysis for eight logarithmic bins in the energy range of 0.3-500\,GeV with $\Gamma$ fixed to the stacking result of $2.32$ (Supplementary Table \ref{table:TS_SED}).
We have tested fixing $\Gamma$ to $2.32$+0.40 and $2.32$-0.35.
The results are all consistent.
The averaged multi-wavelength SED of the FGR sample is shown in Figure \ref{fig:SED_average}.

As a further test,  we stacked the sample from the lowest TS to the highest, and plotted the cumulative TS against the number of sources stacked. 
Supplementary Figure \ref{fig:cumulativeTS} shows the increasing process of the cumulative TS value, indicating that the detected signal is not being dominated by a few bright sources.
In the FGR sample, there are 3 sources with TS value above 8.
As a further test, we stack the 3 sources one by one (Supplementary Figure \ref{fig:profile_onebyone}).
A steady increase of stacked TS is apparent in the TS profiles.
The flux and index in the TS profiles before stacking the 3 sources (Supplementary Figure \ref{fig:profile_onebyone}, panel a), during each individual stacking steps (Supplementary Figure \ref{fig:profile_onebyone}, panel b to e), and the final stacked results of FGR sample (Figure \ref{fig:TS_profile}) are all consistent.
Thus, the 3 sources with comparably high TS values did not dominate the stacking result.

Next, we performed a background stacking analysis\cite{Paliya2019,Ajello2021},
which can test whether the $\gamma$-rays come from the fluctuation of {diffuse} components (GALACTIC and ISOTROPIC). We randomly select 37 empty positions, which are located beyond the 99\% confidence level point source locations of all sources in 4FGL-DR4.
Then the stacking process is repeated again and the stacked TS profile for empty positions are achieved. 
This background stacking analysis was repeated 10 times.
Among the 10 stacked TS profiles for empty positions, the highest stacked TS value is 1.9 as shown in Supplementary Figure \ref{fig:profile_empty}.
There is no significant $\gamma$-ray emission from diffuse component fluctuation, leading to a 95\% confidence level upper limit of $1.79\times10^{-11}\,\fluxunit$. 

We would like to point out how the stacking results depend on the hard X-ray flux cut. 
Considering that the entire sample distribution of the hard X-ray fluxes peaks around $2\times10^{-11}\,{\rm erg\,s^{-1}\,cm^{-2}}$ (Supplementary Figure\,\ref{fig:source}{\textbf a}), we chose it as the cut to build up the control and FGR samples as a natural selection a priori to avoid trails.
On the other hand, we have tested a posteriori that how the significant detection of FGR sample depends on the cut, as shown by Supplementary Figure\,\ref{fig:source}{\textbf b}. 
The relation between TS of FGR sample versus the cuts shows a peak TS value of 34 around $f_{\rm 14-195keV}\sim 1.75\times 10^{-11}\,{\rm erg\,s^{-1}\,cm^{-2}}$.
Therefore, the present results are conservative compared to it.

\subsection{Control sample}
For comparison with our FGR sample, we construct a control sample consisting of $27$ sources from the BASS sample using the same selection criteria outlined in \S\,\ref{sect:sample} 
with the exception of integrated X-ray flux (Supplementary Figure \ref{fig:flowchart}; Supplementary Table \ref{table:control sample}).
As presented in Supplementary Table \ref{table:control sample}, the X-ray fluxes for all sources in the control sample are below $2\times 10^{-11} \,\efluxunit$, which represent a low level of AGN corona activity, opposite to our FGR sample.
The mean distance is $41.7\,(\pm 13.1)$ Mpc, where the uncertainty is the standard deviation of the distribution.
As shown in Supplementary Figure \ref{fig:profile_control}, no significant $\gamma$-ray emission is found.
The maximum TS is only $2.8$ ($1.2\, \sigma$).

\section{Constraints on \texorpdfstring{$\gamma$}{gamma}-rays from AGN corona}
\label{sect:discussion}
It has been suggested that $\gamma$-rays originate from the coronae of AGN, which are radiated by non-thermal electrons\cite{Zdziarski1996} accelerated  either by shocks\cite{Blandford1987, Inoue2019} or magnetic reconnection in hot accretion flows (their Lorentz factors reach even to $\gamma_e\sim 10^6$)\cite{Sironi2014, Chael2017}. These electrons could get energies accompanying formation of the hot corona\cite{Wang2004}.
These energetic electrons are scattering seed photons and generating $\gamma$-rays. 
In the meanwhile, the $\gamma$-rays are not able to escape from the regions if the pair production is optically thick.
The current results provide strong constraints on the spatial distributions of $\gamma$-ray photons.

\subsection{Intrinsic $\gamma$-ray luminosity.}
In this paper, we explore the origination of the observed $\gamma$-ray emissions from non-thermal electrons in the hot corona through inverse Compton scattering of accretion disk emissions as seed photons.
Following the popular model, we assume that the non-thermal electrons with a cutoff power-law spectrum of $n_e(\gamma_e)=n_0\gamma_e^{-p}\exp\left(-\gamma_e/\gamma_{\rm max}\right)$ homogeneously distribute over a sphere corona with radius $R$  after acceleration, where $n_0$ is the normalized number density, $\gamma_e\ge \gamma_{\rm min}$, $\gamma_{\rm min, max}$ are the minimum and  the maximum Lorentz factors, $p$ is the power index.
The electrons are immersed in seed photons with the number density of $n_{\rm ph}(\nu_i)$, where $\nu_i$ is the seed photon frequency. 
We have the $\gamma$-ray spectrum through inverse Compton scattering\cite{Blumenthal1970, Inoue1996},
\begin{equation}\label{eq:Lgamma}
L_{\nu}=8\pi r_e^2hcR^3 \int_{\nu_{\rm min}}^{\nu_{\rm max}}
\int_{\gamma_{\rm min}}^{\infty}
f(x)n_e(\gamma_e)n_{\rm ph}(\nu_i)\,d\gamma_e d\nu_i,
\end{equation}
where $r_e=e^2/m_ec^2$ is the classical electron radius, $h$ is the Planck constant,
$e$ is the electron charge,
$m_e$ is the electron mass, $x=\nu/4\gamma_e^{2}\nu_i$, and the function
\begin{equation}
f(x)=\left\{
\begin{aligned}
&x+2\,x^{2}\,\ln x+x^{2}-2\,x^{3}, & (0<x<1),\\
&0,&(x>1).
\end{aligned}
\right.
\end{equation}
In our calculations, we take $\nu_{\rm min}=10^{13}\,\rm Hz$ and $\nu_{\rm max}=10^{21}\,\rm Hz$ as the minimum and maximum of the incident photon frequencies, respectively, $\gamma_{\rm min}=1$,
$n_{\rm ph}(\nu_i)=L_{\nu_i}/{4\pi R^{2}c h\nu_i}$, and $L_{\nu_i}$ is the specific luminosity derived from the average SED in Figure \ref{fig:SED_average}.

\subsection{Attenuated $\gamma$-rays by pair production.}
The $\gamma$-rays produced in the corona would be attenuated due to pair production ($\gamma\gamma^\prime\rightarrow e^{+} e^{-}$) interactions with the total cross section given by Ref.\cite{Aharonian2004},
\begin{equation}\label{eq:sig_gam}
\sigmagg(\nu,\nu_i)=
\frac{3\,\sigmaT}{2 s^{2}}
\left[\left(s+\frac{1}{2}\ln s-\frac{1}{6}+\frac{1}{2\,s}\right)
\ln\left(\sqrt{s}+\sqrt{s-1}\right)-\right.
\left. \left(s+\frac{4}{9}-\frac{1}{9\,s}\right)\sqrt{1-\frac{1}{s}} \right],
\end{equation}
where $s\equiv h^2\nu \nu_i/m_e^2\,c^{4}$,
$h\nu$ is the energy of the $\gamma$-{rays},
and $h\nu_i$ is the energy of the incident target photon.
$\sigmagg(\nu,\nu_i)$ reaches its maximum value $\approx 0.22 \,\sigmaT$ at $s=3.5$.
For a rough estimation to evaluate the role of pair production, we have
\begin{equation} \label{eq:tau2}
\tau_{\gamma\gamma}\approx\left(\frac{\nu_iL_{\nu_i}}{\Ledd}\right)
\left(\frac{m_{\rm p}c^2}{h\nu_i}\right)
\left[\frac{\sigmagg(\nu,\nu_i)}{\sigmaT}\right]
\left(\frac{R}{\Rg}\right)^{-1} \\
=2.3\left(\frac{\nu_iL_{\nu_i}}{10^{-4}\Ledd}\right)
\left(\frac{R}{10\,\Rg}\right)^{-1},
\end{equation}
showing a key role of the pair production even for AGNs with low Eddington ratios.
The optical depth to pair production is given by
\begin{equation}\label{eq:tau}
\taugg
=\int_{\nu_{\rm min}}^{\nu_{\rm max}}
n_{\rm ph}(\nu_i)\sigmagg(\nu,\nu_i) R d\nu_i.
\end{equation}
Using the soft X-ray Eddington ratio ($\lamX=\Lsoftx/\Ledd\approx \nu_iL_{\nu_i}/\Ledd$) values provided in Supplementary Table \ref{table:faint source},
we find that
$\taugg$ {ranges} from $10^{-1}$ to $10^{3}$,
where $\Lsoftx = 4\pi D^2 F_{\rm pow}$, $D$ is the distance\cite{Koss2022} listed in Supplementary Table \ref{table:faint source}, $F_{\rm pow}$ is the flux of soft X-ray (“PowUnabsFlux” of Table C1 in ref.\cite{Evans2020}), representing the mean total unabsorbed flux assuming a power-law spectrum.
In the uniformity assumption of non-thermal electrons, the $\gamma$-ray photons remaining after attenuation is given by\cite{Dermer2009},
\begin{equation} \label{eq:Lnup}
\Lnup=\frac{3}{\taugg}
\left\{
\frac{1}{2}+\frac{\exp[-\taugg]}{\taugg}-
\frac{1-\exp[-\taugg]}{\taugg^{2}}
\right\} L_{\nu},
\end{equation}
using Equations (\ref{eq:sig_gam}) and (\ref{eq:tau}), as shown in Figure \ref{fig:SED_average}.

In order to estimate the fraction of non-thermal electrons over the thermal emissions of the hot corona, we have $L_{\gamma}(\ge{\rm MeV})=\int_{\rm MeV}L_{\nu}d\nu$, and define a parameter 
\begin{equation} \label{eq:L_gam_int}
\xinth=\frac{L_{\gamma}(\ge\rm MeV)}{\Lx},
\end{equation}
to describe the fraction of non-thermal electrons to the thermal (the X-rays as a proxy for the thermal electron population). In the compact and extended corona, we have $\xicom$ and $\xiext$, respectively.

We would like to note that the accretion shock model is discussed in ref.\cite{Murase2024}, where the primary generation mechanism of $\gamma$-rays is cosmic-ray induced cascade.
Moreover, interactions of proton-proton collisions and proton-photons could also produce $\gamma$-rays in AGNs\cite{Inoue2019}.
The present detections can be applied to constrain these mechanisms in future.
In this paper, the main considered process is the inverse Compton scattering of non-thermal electrons.

\subsection{$\gamma$-ray spectrum fitting for the FGR sample}
As we shown, the hot corona have two regions due to pair production denoted as a compact region $R=R_{\rm com}$ and extended region $R=R_{\rm ext}$. 
We have to calculate $\gamma$-rays from the two regions.
Subsequently, the other parameters ($n_0$ and $p$) will be affiliated with subscripts of ``com" and ``ext" accordingly. We will derive the total number density of non-thermal electrons as
$n_{\rm com}=\int_{{\gamma_{\rm min}}}^{\infty} n_{0,\,\rm com} \gamma_e^{-p_{\rm com}}
{\rm exp}\left(-\gamma_e/\gamma_{\rm max}\right)d\gamma_e$,
$n_{\rm ext}=\int_{{\gamma_{\rm min}}}^{\infty} n_{0,\,\rm ext} \gamma_e^{-p_{\rm ext}}
{\rm exp}\left(-\gamma_e/\gamma_{\rm max}\right)d\gamma_e$ in the two regions from fitting the observed SED, respectively.

The seed photon density can be obtained by a polynomial curve fitting the SED from optical to X-ray band (Figure \ref{fig:SED_average}).
There are three free parameters, $n_{0,\rm com},n_{0,\rm ext},\Rext$ to be determined by the following fitting scheme.
The radius of the compact corona is set to $10\,\Rg$ (ref.\cite{Fabian2015}).
We take the spectrum index of non-thermal electrons $p_{\rm com}=2.9$ obtained from the steady-state solution of the transport equation for the compact corona\cite{Inoue2019}, and
$p_{\rm ext}=1$ for the extended corona\cite{Guo2014} to explain the higher energy $\sim 100\,$GeV photons (SED in Figure \ref{fig:SED_average}).
There is an additional constrain on $\Rext$, the inverse Compton scattering timescale should be comparable to the traveling of the non-thermal electrons, namely
\begin{equation}\label{eq:Rext}
    \frac{\Rext}{c}=\frac{\gamma_e m_ec^2}{P_{\rm IC}},
\end{equation}
where $P_{\rm IC}=2.6\times 10^{-14}\,\gamma_e^2U_{\rm ph}\,{\rm erg\,s^{-1}}$, namely, $\Rext=0.9\,\gamma_5L_{43}\,$pc,
where $\gamma_5=\gamma_e/10^5$, $L_{43}=L_{\gamma\gamma^{\prime}}/10^{43}\,\ergs$ is the $\gamma\gamma^{\prime}$ interaction luminosity.
Otherwise, the extended region of non-thermal electrons cannot be supplied by acceleration.
For a simplified treatment, we assume that the compact and extended corona share the same $\gamma_{\rm max}$ value.

We use Markov chain Monte Carlo (MCMC) algorithm
to fit the $\gamma$-ray spectrum with Equation (\ref{eq:Lnup}) in the parameter space of $n_{0,\rm com},n_{0,\rm ext}$, and $\gamma_{\rm max}$.
Python package $\tt emcee$ is used.
As shown in Supplementary Figure \ref{fig:MCMC}a, the exponential cutoff Lorentz factor $\gamma_{\rm max}$ could not be well constrained.
A conservative value of $\gamma_{\rm max}=10^{5.5}$ is adopted, giving a corresponding radius of $\Rext=2.7\times 10^6\, \Rg$ (or $2.8\,\rm pc$) with Equation (\ref{eq:Rext}).
It is interesting to note that this region is consistent with the regions ($1-23\,$pc) of $\sim100\,$GHz emissions resolved by ALMA observations in radio-quiet AGNs\cite{Ricci2023}. 
Supplementary Figure \ref{fig:MCMC}b gives the best values of $n_{0,\rm ext}$ and $n_{0,\rm com}$.
The number densities of non-thermal electrons in the extended and compact corona are derived as $\Next={6.5\times 10^{-7}}\,\rm cm^{-3}$ and $\Ncom={7.6}\times10^7\,\rm cm^{-3}$, respectively.
The results are shown in Table \ref{table:sed_para}.

\section{Other possible origins of the \texorpdfstring{$\gamma$}{gamma}-ray emission {from the} FGR sample}
\label{sect:other_origin}
\subsection{$\gamma$-ray contribution due to star formation.}

Since star formation can produce $\gamma$-ray photons due to interactions between high-energy cosmic rays and the interstellar medium,
we have excluded star forming galaxies from the FGR sample and control sample to minimize their contribution.
For a further test, we discuss the $\gamma$-ray contribution of star formation activity in the FGR sample, determined based on the far-infrared luminosity.
If we assume the infrared luminosities of the FGR sample are dominated by star formation activity,
we could estimate the star-formation contribution to the $\gamma$-ray luminosity using the $L_{0.1-800\,\rm GeV}-\LIR$ relation established by ref.\cite{Ajello2020} for resolved and unresolved galaxies combined.
\begin{equation}
\label{eq:Lgam-LIR}
\log\left(\frac{L_{0.1-800\,\rm GeV}}{\rm erg \,s^{-1}}\right)=(39.20^{+0.06}_{-0.05})
+(1.15^{+0.08}_{-0.03})\log \left(\frac{\LIR}{10^{10}\,\Lsun}\right).
\end{equation}
We obtain the average $\log(\LIR / \Lsun)={10.12}$ using the average SED in Figure \ref{fig:SED_average},
corresponding to a $\gamma$-ray luminosity of {$L_{\gamma,\rm\,SFG}={2.16}\times 10^{39}\rm erg \,s^{-1}$} in the 0.1-800\,GeV energy band.
This value is only around a sixteenth of the stacking luminosity $L_{\rm 0.1-800\,GeV} =(3.6\pm 1.7)\times10^{40}{\,\rm erg\,s^{-1}}$ of FGR sample, derived from $\LGam$ with the fitted index of $\Gamma=2.32$. 
The average infrared luminosity of the control sample is $\log(\LIR / \Lsun)={10.07}$ based on their average SED.
Adopting the above $L_{0.1-800\,\rm GeV}-\LIR$ relation, the corresponding $\gamma$-ray luminosity is $L_{\gamma,\rm\,SFG}={1.91}\times 10^{39}\rm erg \,s^{-1}$ in the 0.1-800\,GeV energy band,
which is consistent with the FGR sample and no significant $\gamma$-ray emission is detected.
Hence, we conclude that contribution from star formation is negligible.

\subsection{$\gamma$-rays contributed from low-power jets.}

The $\gamma$-ray emission we observed from the FGR sample is unlikely to be produced by low-power jets.
For the FGR sample, its $\LnuI/\Lx$ follows a $\sim10^{-5}$ on average,
which indicates that the radio emission is mainly contributed by the hot AGN corona\cite{Laor2008, Smith2020, Ajello2021}.
More importantly, 
the average 3 GHz radio luminosity for the control sample is $\log( \LnuIII/{\rm erg \,s^{-1}}) = 37.5\,(\pm 0.5)$, similar with our FGR sample, $\log( \LnuIII/{\rm erg \,s^{-1}})=37.4\,(\pm 0.9)$, but no significant $\gamma$-ray emission is detected,
where the errors are given by the standard deviation.
Therefore, we conclude that low-power jets in the FGR sample cannot explain their stacked $\gamma$-ray emission.

Additionally, the averaged GHz radio luminosity of the FGR sample is $\sim$10$^{38}$ erg s$^{-1}$ (Figure \ref{fig:SED_average}). 
For the self-synchrotron Compton (SSC) model of jets, we have to assume the equipartition between the magnetic field ($U_{\rm B}$) and seed photon energy density ($U_{\rm syn}$) to avoid the so-called Compton catastrophe. In the SSC model, the $\gamma$-ray luminosities produced by the jet through inverse Compton scattering should also be $\sim$10$^{38}$ erg s$^{-1}$ since $L_{\rm IC}\approx (U_{\rm syn}/U_{\rm B})L_{\rm syn}\approx L_{\rm syn}$, if we assume that all radio emissions originate from the jet. 
This $\gamma$-ray luminosity is about two orders of magnitude lower than the stacking result of $\sim$10$^{40}$ erg s$^{-1}$.
Thus, even if we assume all the radio emissions of FGR sample are contributed by jets, their contribution of $\gamma$-rays can be ignored comparing to AGN corona.
The population of electrons radiating radio emissions are poorly constrained by the current data, such as the maximum Lorentz factor and energy distribution, since near infrared continuum is fully dominated by the thermal emissions from the torus.
This makes it hard to calculate its $\gamma$-ray contribution to the observed {\textit{Fermi}}-LAT spectra under other radiation process beyond SSC (e.g. external inverse Compton scattering).
However, the low power jets are not expected to contain very energetic electrons to radiate very high $\gamma$-ray flux in radio-quiet AGNs\cite{Laor2008}.
Hence even in the context of seed photon density $U_{\rm ph}$ much higher than $U_B$, the low power jets through external inverse Compton scattering are not likely able to produce the $\sim$GeV photons presented in this paper.

There is a possibility that the jets are bended, and we lie within the $\gamma$-ray beaming cone but not the radio beaming cone. 
Such sources would appear as $\gamma$-ray loud but radio-quiet AGNs to us, similar as FGR sample, indicating that jet bending plays an important role in gamma-ray detection of AGNs.
However, it is reasonable to assume that substantial radio emission occurs in regions of gamma-ray emission, making this scenario unlikely\cite{Graham2014}.
Additionally, ref.\cite{Graham2014} shows that jet bending is not a significant factor for gamma-ray detection in AGNs.
Thus, it is unlikely that the stacked gamma-ray emission we observed from FGR sample originates from jets.

\subsection{$\gamma$-ray contribution from outflow.}
Recent work about $\gamma$-ray emission from galaxies hosting molecular outflow shows no evidence that the outflows are accelerating charged particles directly, {but they may produce more $\gamma$-rays than galaxies without outflows\cite{McDaniel2023}}.
A stacked analysis of galaxies with a highly ionized ultrafast ($v>0.1\,c$) outflow (UFO) revealed the detection of significant $\gamma$-ray emission\cite{Ajello2021}.
However, the ionized outflow velocities of sources in BASS sample are generally smaller than $0.01\,c$ (see ref.\cite{Rojas2020}).
Our FGR sample does not overlap with the sample in ref.\cite{Ajello2021}.
Weak undetected UFOs may exist in FGR sample but their gamma-ray emission should be minimal\cite{Ajello2021}.
Therefore, the $\gamma$-ray contribution from molecular {outflows} or {UFOs} is negligible in this work.

\section{Results of NGC 3281}
\label{sect:NGC result}
We {analyzed} {{\it Fermi}-LAT} data for NGC 3281.
Assuming a power-law model, it is detected with a TS value of  $46.5$ and a photon index of $2.63(\pm0.02)$ in 0.1-300\,GeV. The $\gamma$-ray position of NGC 3281 is consistent with the optical position (see Supplementary Figure \ref{fig:TSmap_NGC}).
Although the TS value of NGC 3281 is larger than 25, it {is located} near a region of extended residual {emission}, {which brings uncertainty} to its detection.
Future {observations} with {{\it Fermi}-LAT} will clarify on this point.
We exclude NGC 3281 from further study in this paper.

\section{Non-blazars in the BASS-4FGL sample}
\label{sect:cross match}

Here we discuss the sources that arise from BASS $\&$ 4FGL catalogs cross-matching (BASS-4FGL non-blazar sample hereafter; see \S\,\ref{sect:sample}).
The BASS-4FGL non-blazar sample contains 9 sources.
{Besides corona, their $\gamma$-ray emission may have other origins}.
Among the 9 sources in the BASS-4FGL non-blazars sample,
Circinus Galaxy is a nearby edge-on spiral starburst galaxy\cite{Freeman1977}. 
10 years of {\it Fermi}-LAT data were analyzed\cite{Guo2019}, indicating that Circinus is a composite starburst-AGN system.
The potential contribution of corona and accretion shock to its sub-GeV $\gamma$-ray emission was recently discussed\cite{Murase2024}.

For the 8 remaining sources, their $\gamma$-ray localizations (see Supplementary Figure \ref{fig:TSmap_non-blazar}) {cover other possible $\gamma$-ray emitting counterparts,} contaminating any $\gamma$-rays that could be attributed to coronae.
For NGC 4151, ESO 354-4, ESO 253-G003, Mrk 520, and 2MASX J09023729-4813339, their $\gamma$-ray {source positions are also consistent with} blazars 1E 1207.9+3945\cite{Ajello2021, Murase2024}, PMN J0151-3605, PKS 0524-460, TXS 2157+102, and PMN J0903-4805, respectively\cite{Ballet2023}.
Nevertheless, recent studies of NGC 4151 have raised the possibility that the $\gamma$-ray emission may originate from ultra-fast outflows\cite{Peretti2023} or activity in the jet or the corona\cite{Inoue2023}.
For LEDA 154696, the association suggested in the 4FGL-DR4 catalog is CRATES J100710-095715, a flat spectrum radio source\cite{Healey2007}.
But currently it is {outside} of the positional error in our analysis result.
WISEA J100714.48-094902.1, a source from the WIBRaLS {catalog, is spatially consistent with the $\gamma$-ray emission}.
The $\gamma$-ray emission spatially associated with HE 0436-4717 also covers pulsar PSR J0437-4715 (see ref.\cite{Ballet2023}).
Our localization for LEDA 50427 deviates from the position of association in 4FGL-DR4 catalog by $\sim0.3^{\circ}$.
We searched the vicinity of the new localization and found that TXS 1404-300, a radio source, is spatially consistent.

\clearpage
\begin{addendum}
\item[Data availability]
The {\it Fermi}-LAT data are publicly available at: https://fermi.gsfc.nasa.gov/ssc/data/access/.

\item The {{\it Fermi}-LAT} Collaboration acknowledges support for LAT development, operation and data analysis from NASA and DOE (United States), CEA/Irfu and IN2P3/CNRS (France), ASI and INFN (Italy), MEXT, KEK, and JAXA (Japan), and the K.A.~Wallenberg Foundation, the Swedish Research Council and the National Space Board (Sweden). Science analysis support in the operations phase from INAF (Italy) and CNES (France) is also gratefully acknowledged. This work performed in part under DOE Contract DE-AC02-76SF00515.
Useful discussions are acknowledged with P. Du, Y.-R. Li, Y.-J. Chen and Y.-L. Wang from IHEP AGN Group. We thank the support from NSFC(-12333003, -12273038, -11991050, -11991054), from the National Key R\&D Program of China (2020YFC2201400, 2021YFA1600404).

\item[Author Contributions]
JMW conceived the project of $\gamma$-rays from radio-quiet AGNs and suggested the current corona model. JL led the project and drafted the first version of the paper. JRL led reduction of the {\it Fermi} data and drafted the Materials and Methods. JL and JMW led the revision of the manuscript by considering suggestions and comments from all the authors. All the authors discussed the contents, and revised first version and form the final version of the paper.

\item[Competing interests]
The authors declare no competing interests.

\end{addendum}

\clearpage

\begin{table*}[htbp!]
\centering
\begin{threeparttable}
\footnotesize 
\caption{\bf The parameters of the hot corona through fitting {\it Fermi}-LAT observations}
\begin{tabular}{lcccc cccc}
\hline\hline
Name& $n_{0,\rm \,com}$ & $n_{0,\rm\, ext}$ & & $\Rcom$ & $\Rext$ &$\gamma_{\rm max}$\\
& [$\rm cm^{-3}$] & [$\rm cm^{-3}$] & & [$\Rg$] & [$\Rg$]\\
\hline
FGR sources & $5.4\times10^{-8}$ &$1.4\times 10^8$&&$10$&$2.7\times 10^6$& $10^{5.5}$  \\
\hline
\end{tabular}
\begin{tablenotes}[flushleft]
\footnotesize 
\item {\rmfamily{\bf Notes.} $n_{0,\rm \,com}$, $n_{0,\rm\, ext}$, $\Rext$, and $\gamma_{\rm max}$ are derived from MCMC simulation and $\Rcom$ is fixed at $10\,\Rg$.}
\end{tablenotes}
\label{table:sed_para}
\end{threeparttable}
\end{table*}

\clearpage

\clearpage

\clearpage




\renewcommand{\figurename}{Supplementary Figure} 
\renewcommand{\tablename}{Supplementary Table} 
\setcounter{figure}{0}
\setcounter{table}{0}

\clearpage
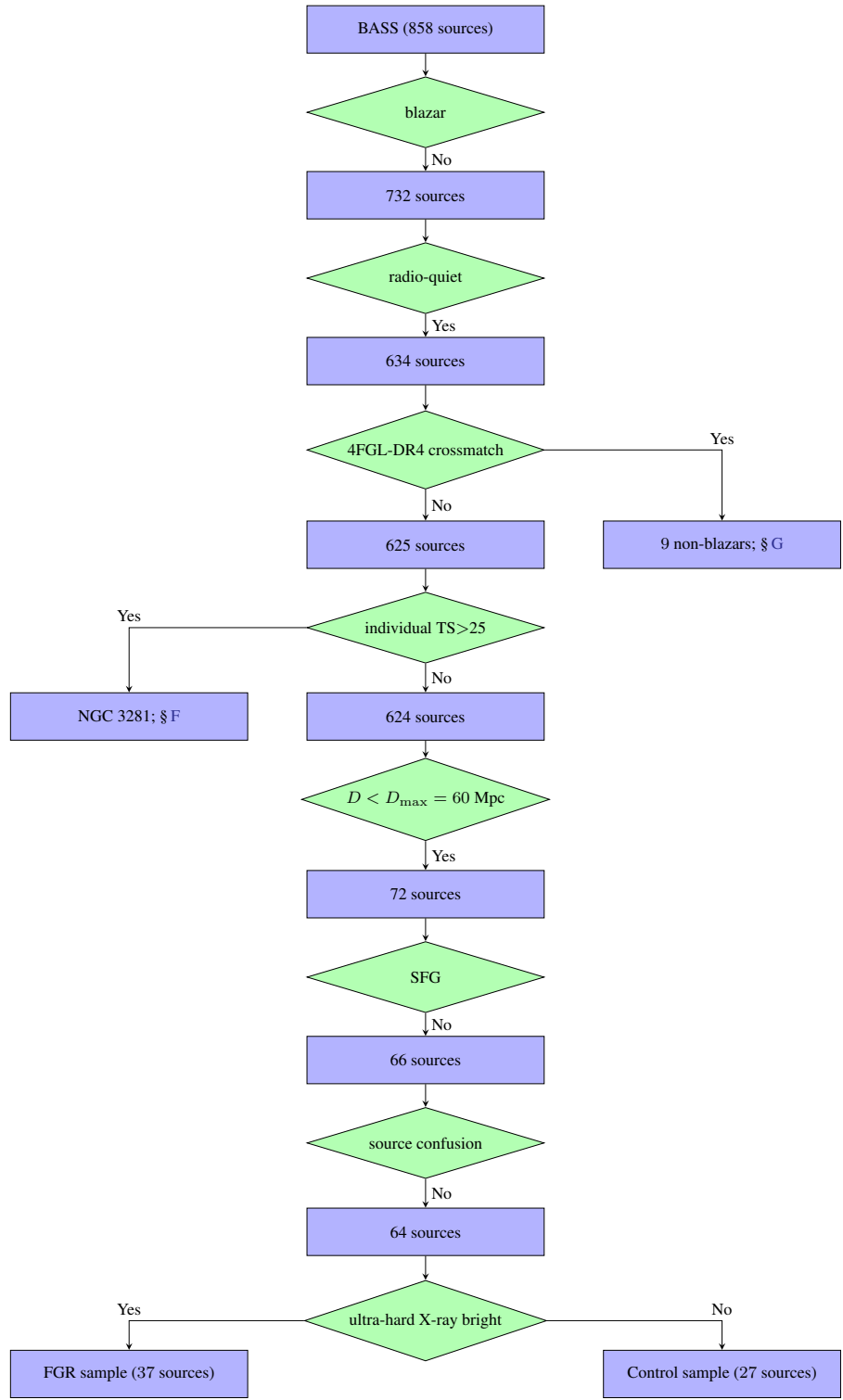
\begin{figure*}
\footnotesize
\centering
\tikzstyle{startstop} = [rectangle, rounded corners, minimum width = 2cm, minimum height=1cm,text centered, draw = black, fill=red!30]
\tikzstyle{io} = [rectangle, minimum width=4cm, minimum height=0.8cm, text centered, draw=black, fill=blue!30]
\tikzstyle{decision} = [diamond, aspect = 3, text centered, draw=black, minimum width=4cm, minimum height=1.2cm,fill=green!30]
\tikzstyle{arrow} = [->,>=stealth]
\scalebox{0.84}{
\begin{tikzpicture}
\centering
\node[io, align=center](858src){BASS (858 sources)};
\node[decision, below of = 858src, yshift = -0.4cm, xshift=0cm, align=center](nonblz){{blazar}};
\node[io, below of = nonblz, yshift = -0.4cm, xshift=0cm](732src){{732 sources}};
\node[decision, below of = 732src, yshift = -0.4cm, xshift=0cm, align=center](radioquiet){radio-quiet};
\node[io, below of = radioquiet, yshift = -0.4cm, xshift=0cm](634src){634 sources};
\node[decision, below of = 634src, yshift = -0.5cm, xshift=0cm](separation){4FGL-DR4 crossmatch};
\node[io, below of = separation, yshift = -0.6cm, xshift=5cm](9src){9 non-blazars};
\node[io, below of = separation, yshift = -0.6cm, xshift=0cm](625src){625 sources};
\node[decision, below of = 625src, yshift = -0.4cm, xshift=0cm](TS25){individual TS$>$25};
\node[io, below of = TS25, yshift = -0.5cm, xshift=0cm](624src){624 sources};
\node[io, below of = TS25, yshift = -0.5cm, xshift=-5cm](1src){NGC 3281};
\node[decision, below of = 624src, yshift = -0.4cm, xshift=0cm, align=center](nearby){$D<D_{\rm max}=60$ Mpc};
\node[io, below of = nearby, yshift = -0.6cm](72src){72 sources};
\node[decision, below of = 72src, yshift = -0.4cm, xshift=0cm, align=center](SFG){{SFG}};
\node[io, below of = SFG, yshift = -0.4cm](66src){66 sources};
\node[decision, below of = 66src, yshift = -0.4cm, xshift=0cm, align=center](hardX){{source confusion}};
\node[io, below of = hardX, yshift = -0.5cm](64src){64 sources};
\node[decision, below of = 64src, yshift = -0.5cm, xshift=0cm](Xray){ultra-hard X-ray bright};
\node[io, below of = Xray, yshift = 0.1cm, xshift=-5cm](FGR){FGR sample (37 sources)};
\node[io, below of = Xray, yshift = 0.1cm, xshift=5cm, align=center](control){Control sample ($27$ sources)};
\draw [arrow] (858src) -- (nonblz);
\draw [arrow] (nonblz) -- node [right] {No} (732src);
\draw [arrow] (732src) -- (radioquiet);
\draw [arrow] (radioquiet) -- node [right] {Yes} (634src);
\draw [arrow] (634src) -- (separation);
\draw [arrow] (separation) -- node [right] {No} (625src) ;
\draw [arrow] (separation) -| node [above] {Yes} (9src);
\draw [arrow] (625src) -- (TS25);
\draw [arrow] (TS25) -- node [right] {No} (624src);
\draw [arrow] (TS25) -| node [above] {Yes} (1src) ;
\draw [arrow] (624src) -- (nearby);
\draw [arrow] (nearby) -- node [right] {Yes} (72src);
\draw [arrow] (72src) -- (SFG);
\draw [arrow] (SFG) --node [right] {No}  (66src);
\draw [arrow] (66src) -- (hardX);
\draw [arrow] (hardX) -- node [right] {No} (64src);
\draw [arrow] (64src) -- (Xray);
\draw [arrow] (Xray) -| node [above] {Yes} (FGR);
\draw [arrow] (Xray) -| node [above] {No} (control);
\end{tikzpicture}}
\caption{\footnotesize Flow chart of sample selection.
From the initial BASS sample, 37 sources are selected as the FGR sample and 27 sources are selected as the control sample.}
\label{fig:flowchart}
\end{figure*}

\clearpage
\begin{figure*}
\centering
\includegraphics[scale=0.9,trim=0 0 0 0]{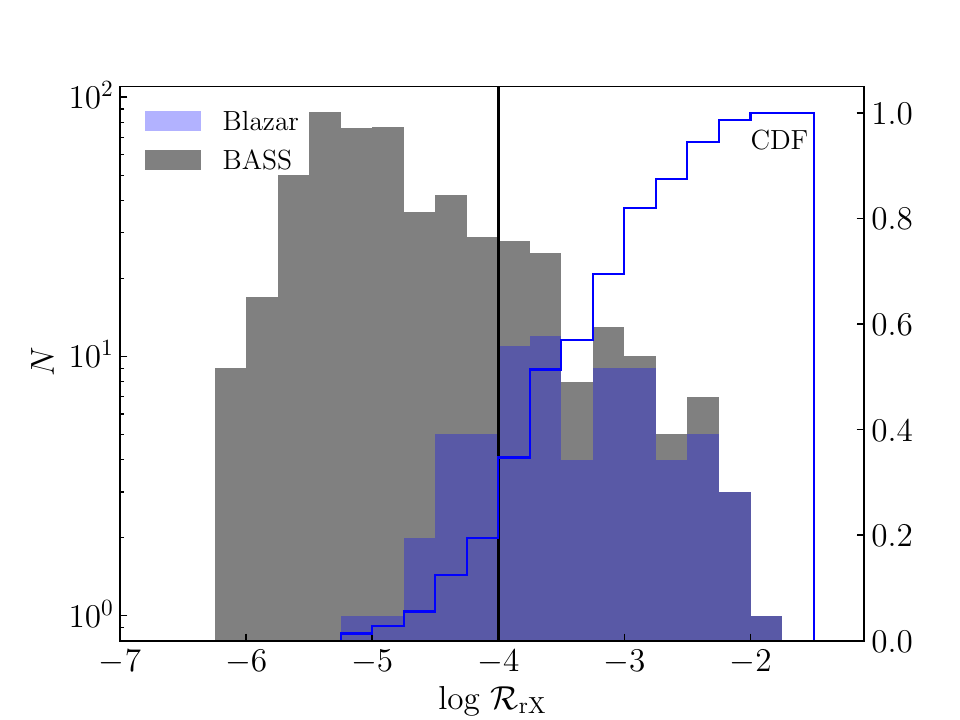}
\caption{\footnotesize
Histogram of the hard X-ray radio-loudness parameter values for BASS sample.
The gray histogram represents the $\Rrx$ distribution of the whole BASS sample.
The light-blue histogram represents the $\Rrx$ distribution of blazars in the BASS sample.
The blue solid line represents
the normalized cumulative distribution function (CDF) of blazars in the BASS sample.
The black vertical line represents the critical value $\Rrx=10^{-4}$.}
\label{fig:histRrx}
\end{figure*}

\begin{figure*}
\centering
\includegraphics[scale=0.5]{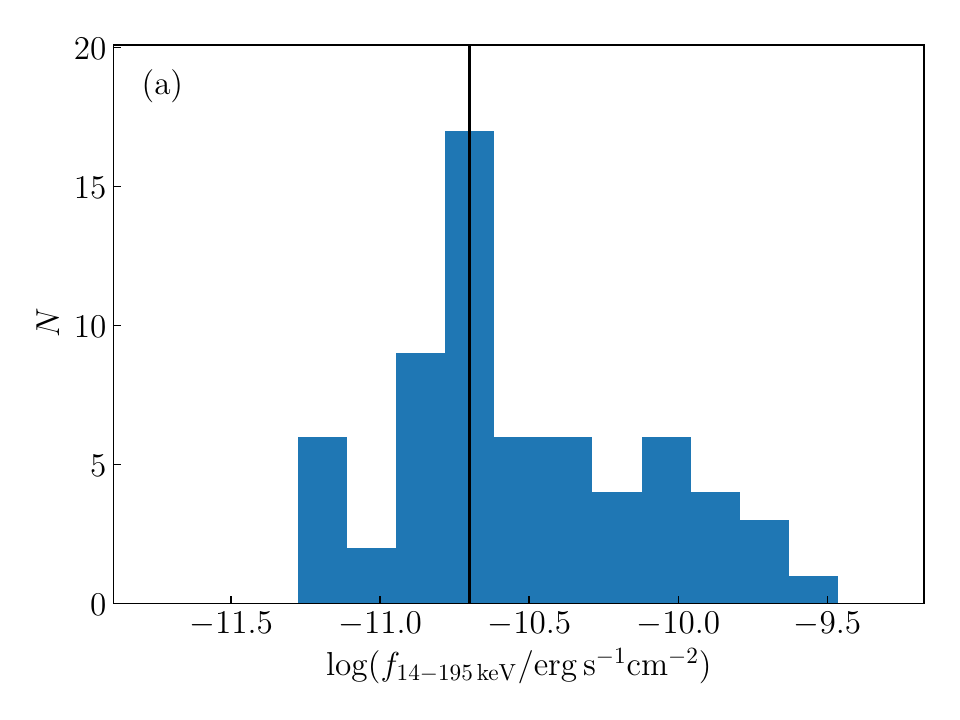}%
\includegraphics[scale=0.5]{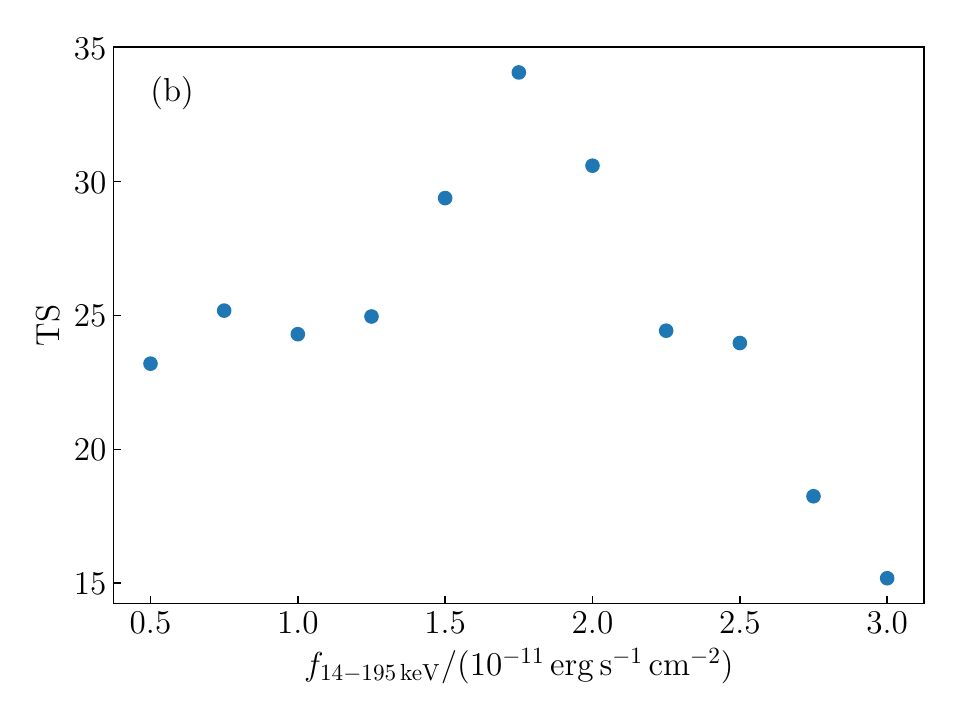}
\caption{
Source distribution against the X-ray flux, and the stacked TS values against the X-ray flux cut.
Panel (a): source distribution against X-ray flux in the 14-195 keV of FGR sample and control sample combined.
The black line indicates the X-ray flux cut as $2\times 10^{-11} \,\efluxunit$ adopted in this paper.
Panel (b): the stacked TS values of the FGR sample against the X-ray flux cut.
}
\label{fig:source}
\end{figure*}

\clearpage
\begin{figure*}
\centering
\includegraphics[scale=0.9,trim= 0 0 0 0]{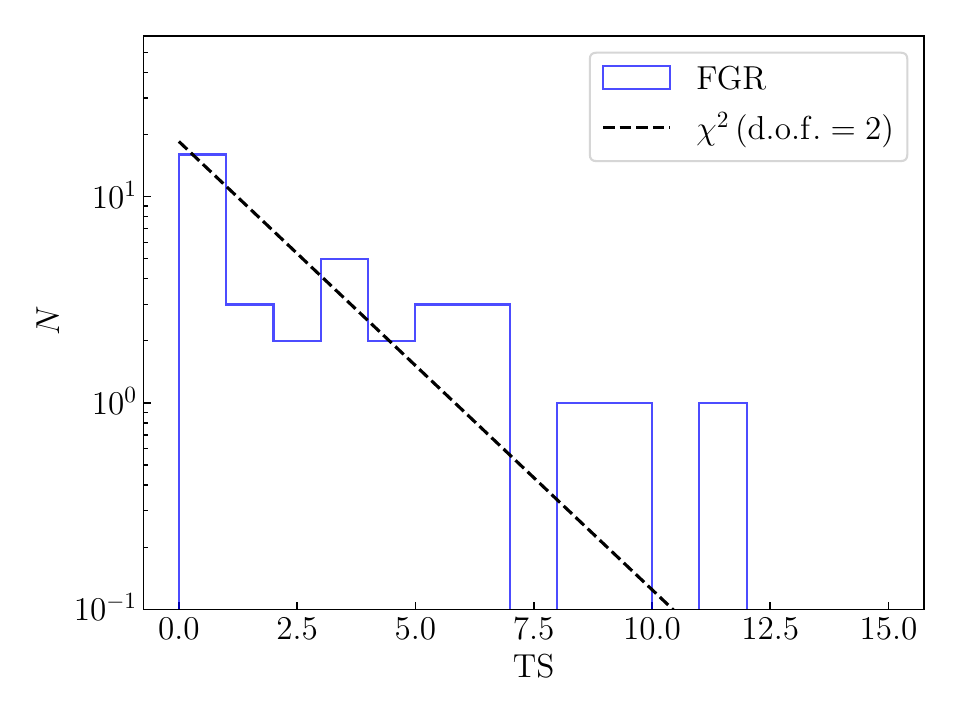}
\caption{\footnotesize
TS distribution of the FGR sample. {The black} dashed line {represents} the TS fluctuations induced by the diffuse $\gamma$-ray background,
which is shown as $\chi^2(\rm d.o.f.=2)$ distribution. }.
\label{fig:TS_distribution}
\end{figure*}


\clearpage
\begin{figure*}
\centering
\includegraphics[scale=0.9]{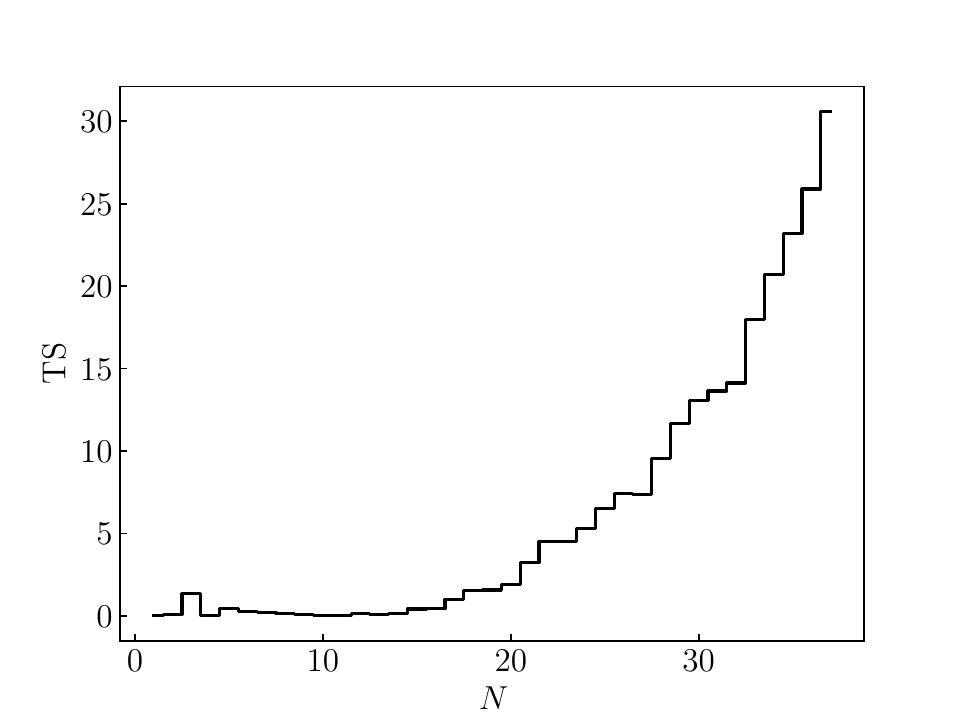}
\caption{\footnotesize
The cumulative TS value versus the number of sources stacked.
The FGR sample are stacked from the lowest TS to the highest one by one.}
\label{fig:cumulativeTS}
\end{figure*}

\clearpage
\begin{figure*}
\centering
\includegraphics[scale=0.48, trim=0 40 0 30]{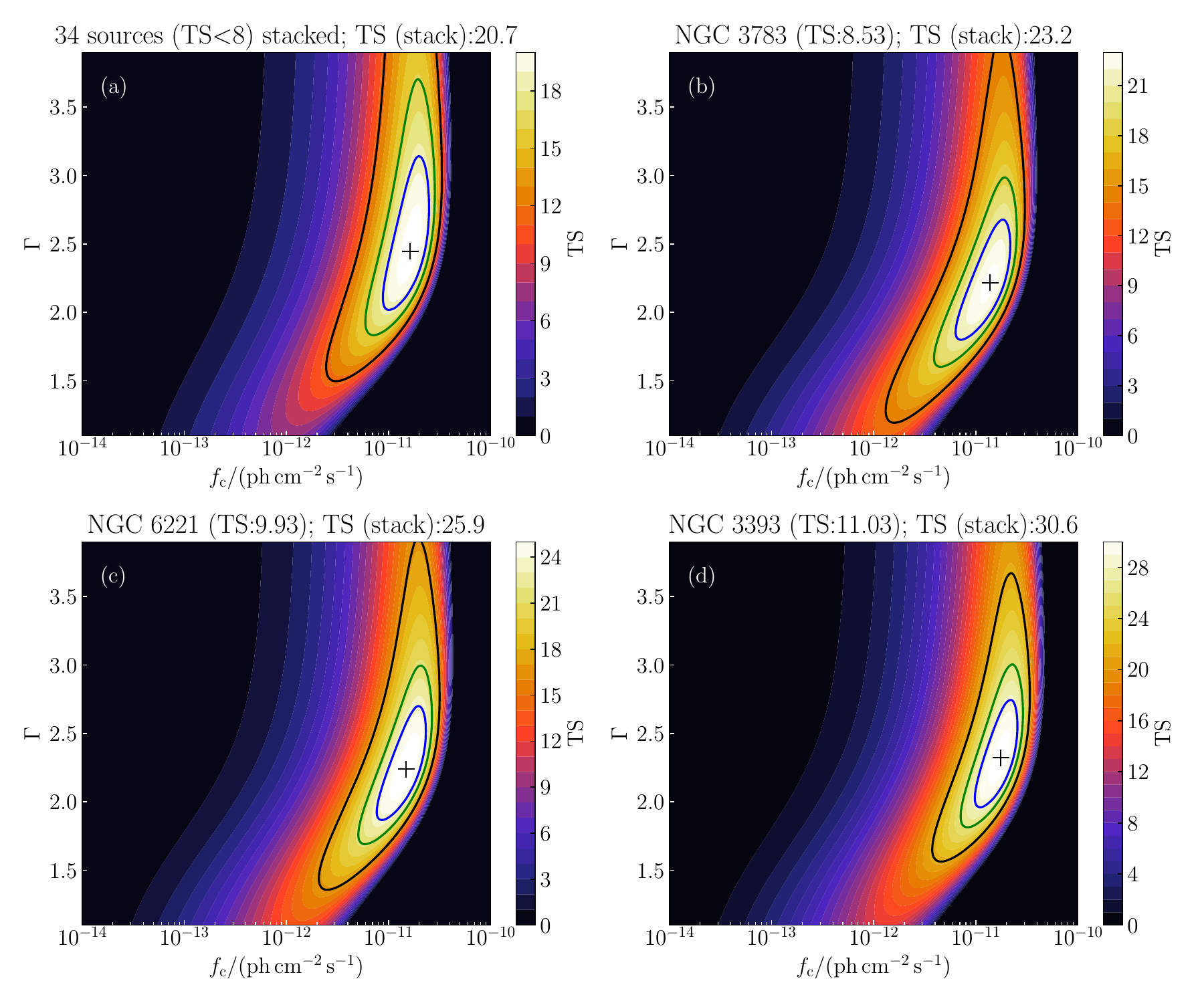}
\caption{\footnotesize
TS profiles of FGR sample in stacking analysis.
Panel (a): the stacked TS profile of 34 sources (TS $<$ 8) in FGR sample.
Panel (b) to (d): the TS profiles after stacking the 3 sources (TS $>$ 8) one by one.
The best fit result is marked by the black cross. 
Three solid contours present the 68\%, 90\% and $ 99\% $ confidence levels.
}
\label{fig:profile_onebyone}
\end{figure*}

\clearpage
\begin{figure*}
\centering
\includegraphics[width=\textwidth, trim=0 0 0 0]{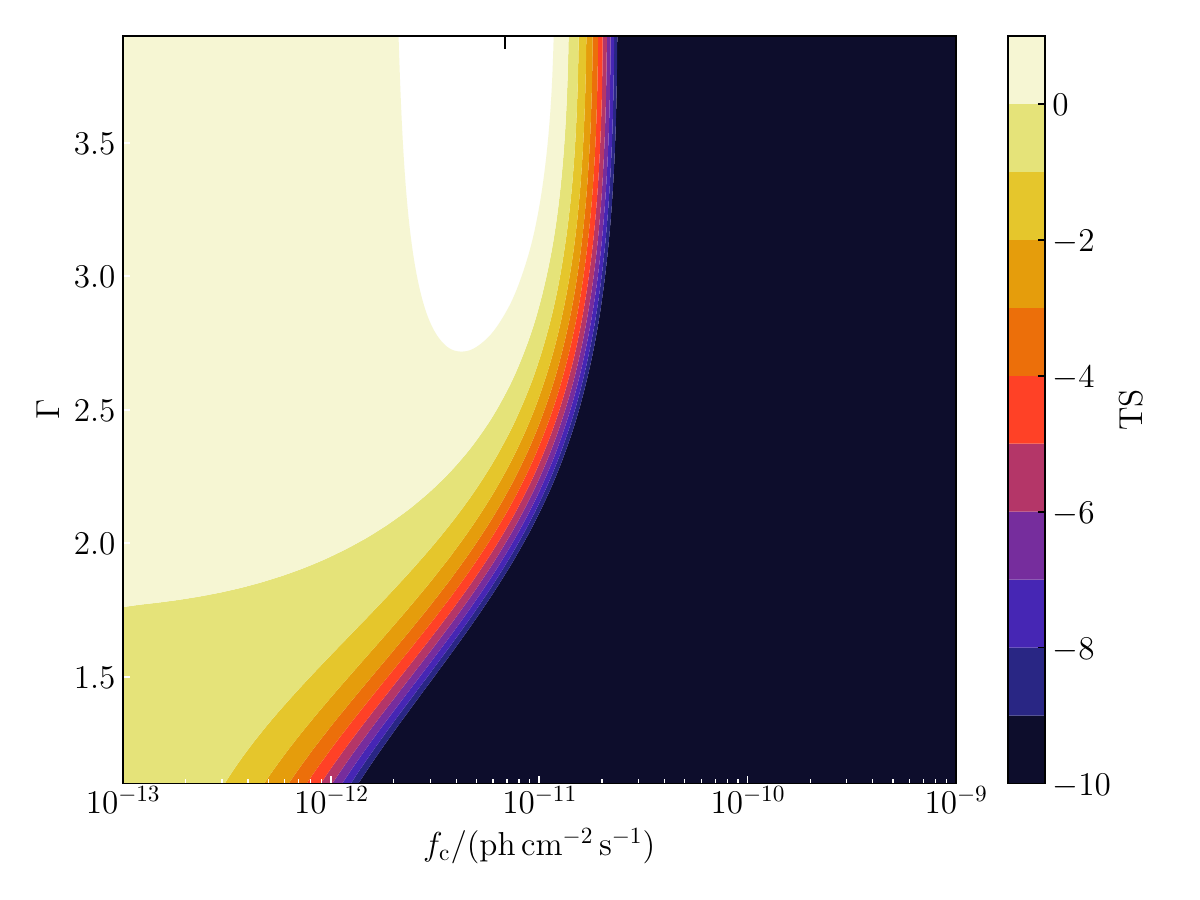}
\caption{\footnotesize
Stacked TS profile of random empty positions.
TS profiles of 37 random empty positions are stacked,
which are far away from the bright $\gamma$-ray sources in the 4FGL-DR4 catalog. The maximum TS in the profile is smaller than 2 and therefore there is no obvious peak.
}
\label{fig:profile_empty}
\end{figure*}

\clearpage
\begin{figure*}
\centering
\includegraphics[width=\textwidth, trim=0 0 0 0]{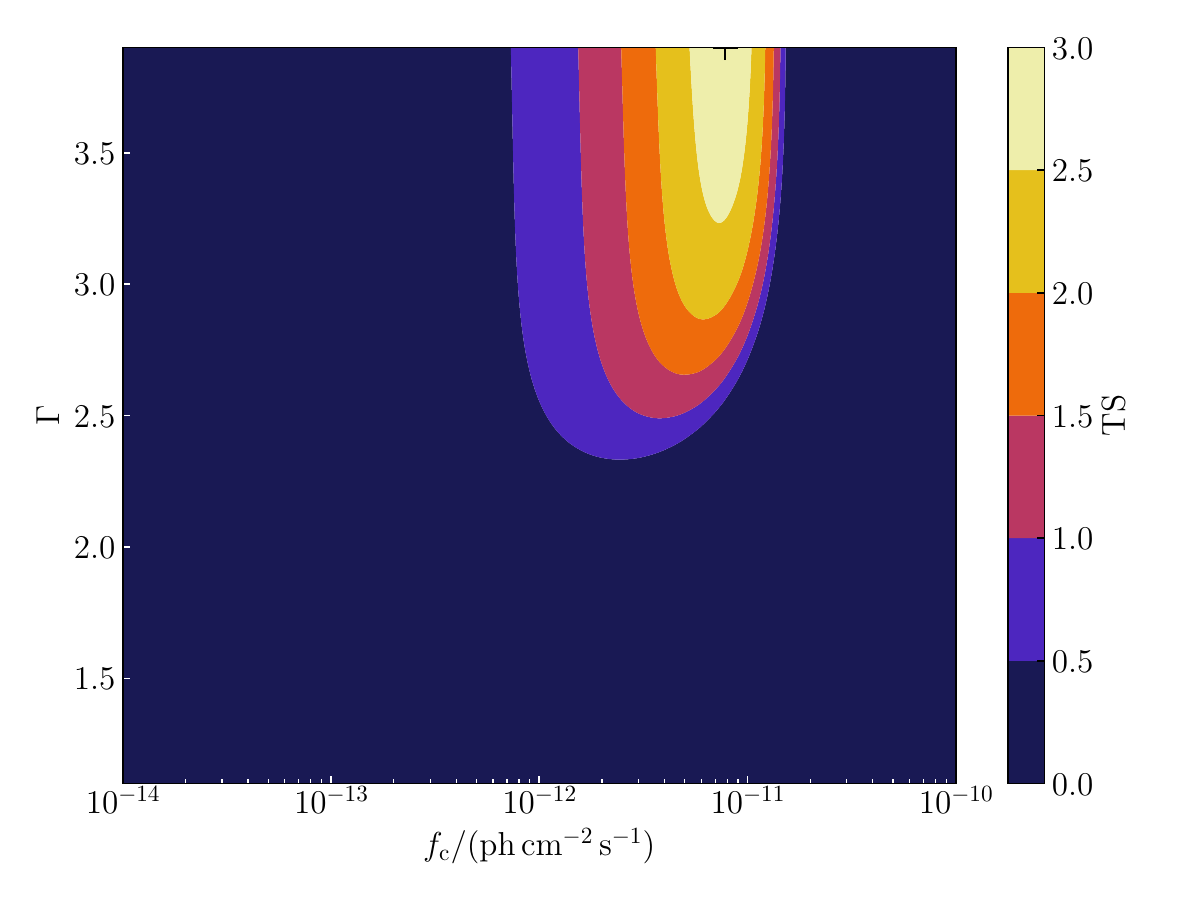}
\caption{\footnotesize
Stacked TS profile of the control sample.
TS profiles of the $27$ sources in the control sample are stacked. The maximum TS value of which is $2.8$ ($1.2\, \sigma$).
}
\label{fig:profile_control}
\end{figure*}

\clearpage
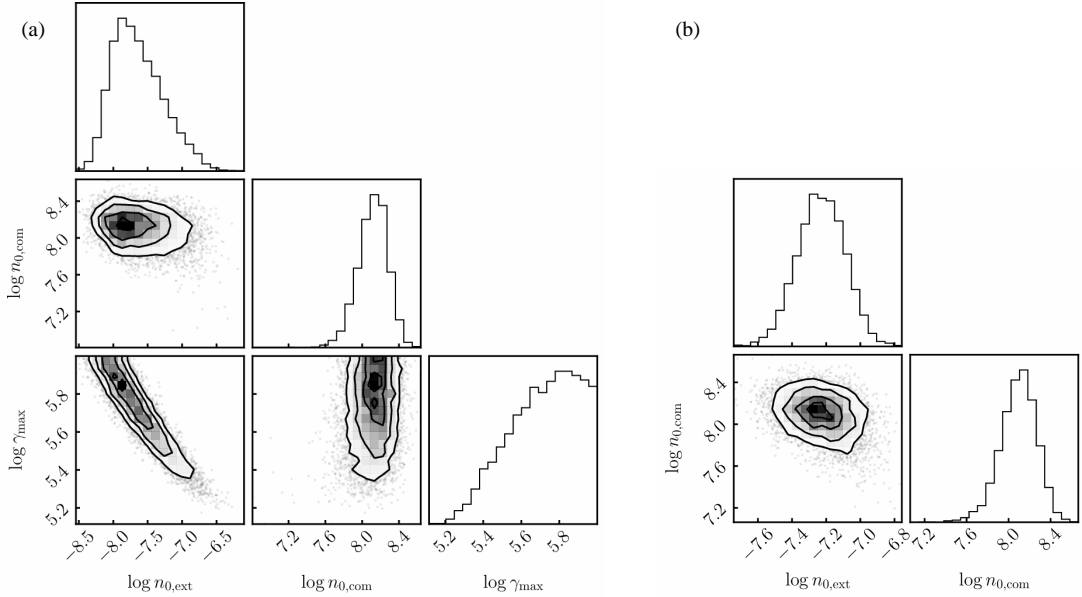
\begin{figure*}
\centering
\begin{subfigure}[b]{0.53\textwidth}
\centering
\includegraphics[width=\textwidth]{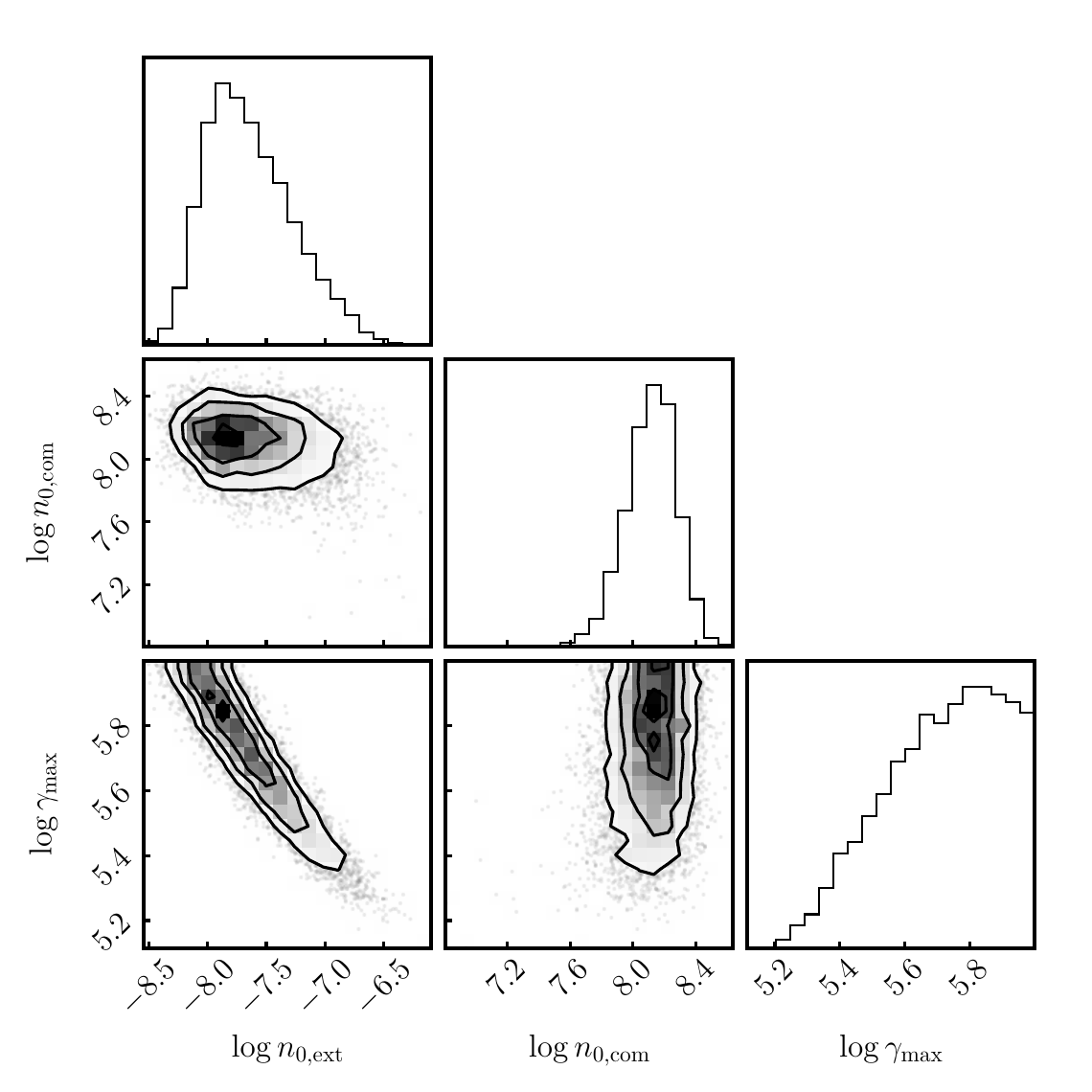} 
\begin{tikzpicture}[overlay, remember picture]
    \node at (-4, 9.5) {\footnotesize (a)};
\end{tikzpicture}
\end{subfigure}
\hspace{-0.5cm} 
\vspace{-1.5cm} 
\begin{subfigure}[b]{0.38\textwidth}
\centering
\includegraphics[width=\textwidth]{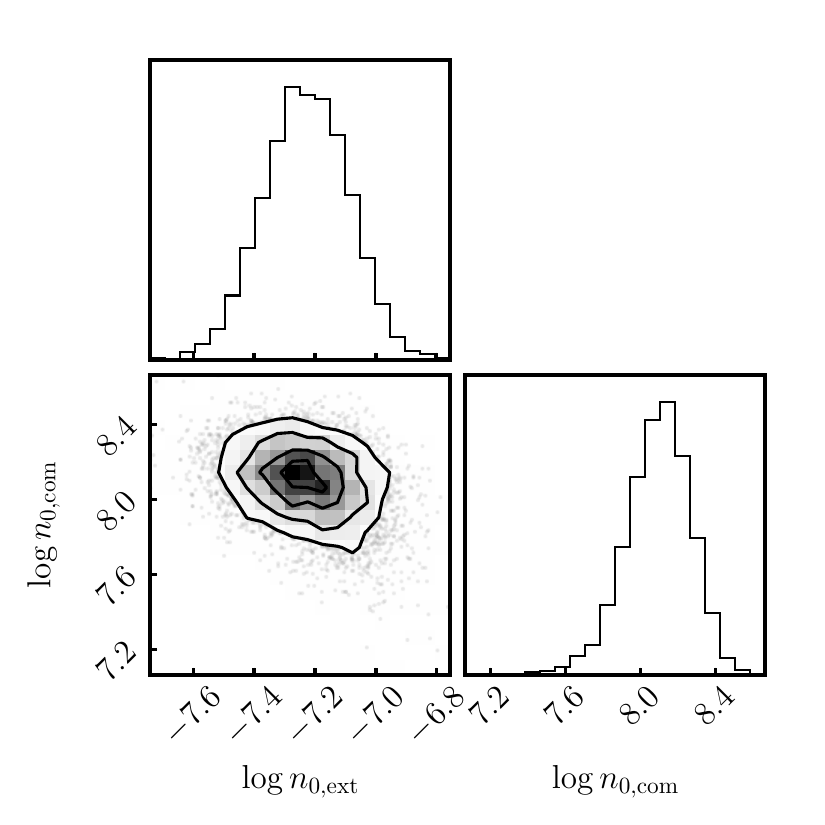} 
\begin{tikzpicture}[overlay, remember picture]
    \node at (-3, 9.5) {\footnotesize (b)};
\end{tikzpicture}
\end{subfigure}
\caption{\footnotesize
Fitting results of {\textit{Fermi}}-LAT SED through MCMC with the normalized number density and the maximum Lorentz factor of the non-thermal electrons.
Panel (a): Three parameters ($n_{0,\rm ext}$, $n_{0,\rm com}$, and $\gamma_{\rm max}$) are fitted.
Panel (b): 
Parameters $n_{0,\rm ext}$ and $n_{0,\rm com}$ are fitted for fixed $\gamma_{\rm max}=10^{5.5}$.}
\label{fig:MCMC}
\end{figure*}

\clearpage
\begin{figure*}
\centering
\includegraphics[scale=0.7,trim= 0 100 0 100]{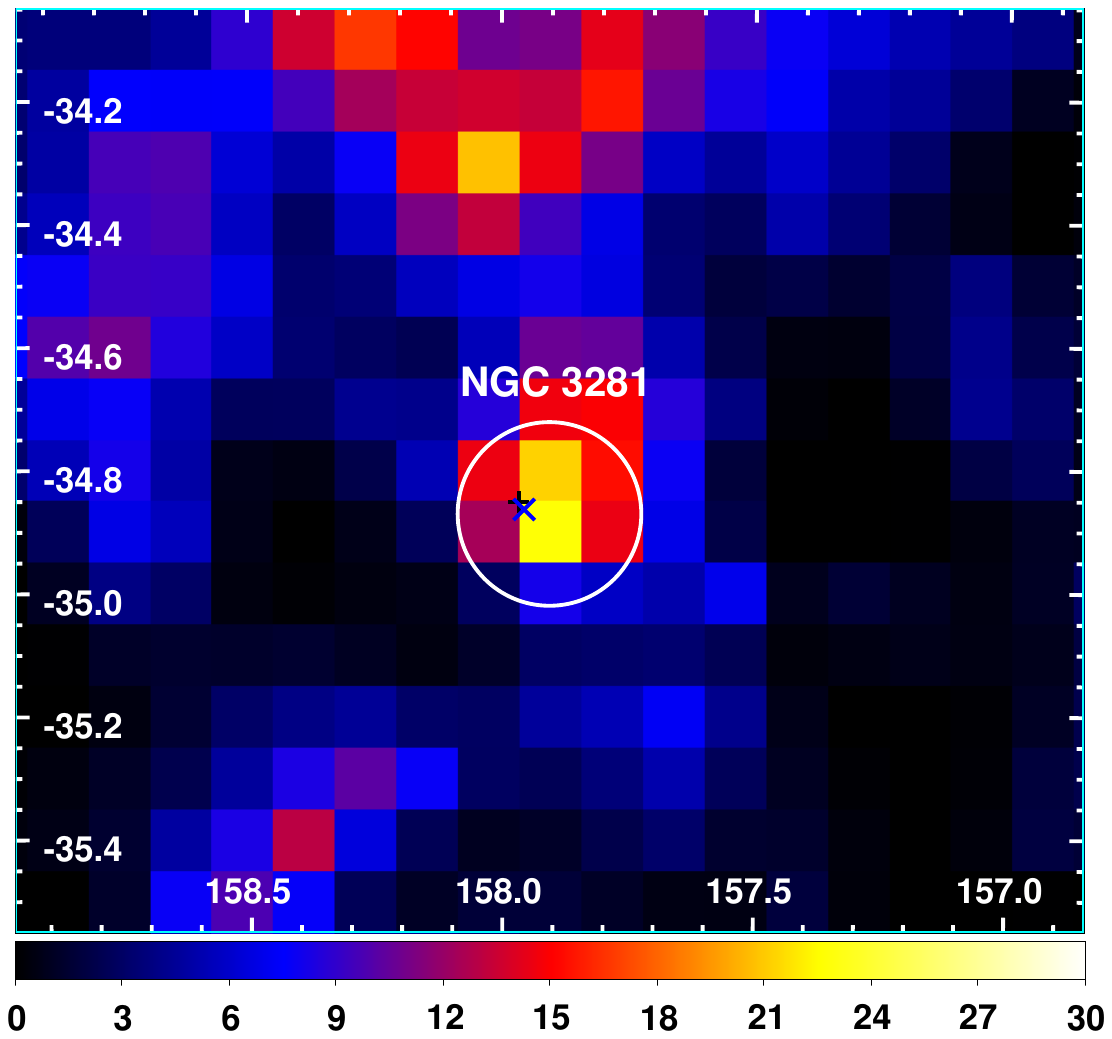}%
\caption{\footnotesize
TS map of NGC 3281 in 0.1-300\,GeV. The black cross indicates the position of the optical source. 
The blue X indicates the position of the associated X-ray source. 
The white circle indicates the 95$\%$ confidence error circle of the $\gamma$-ray localization. The color bar indicates the range of TS values.
The $x$ and $y$ axes are R.A. and decl. (J2000) in degrees.}
\label{fig:TSmap_NGC}
\end{figure*}

\clearpage
\begin{figure*}
\centering
\includegraphics[scale=0.5, trim=80 120 0 150]{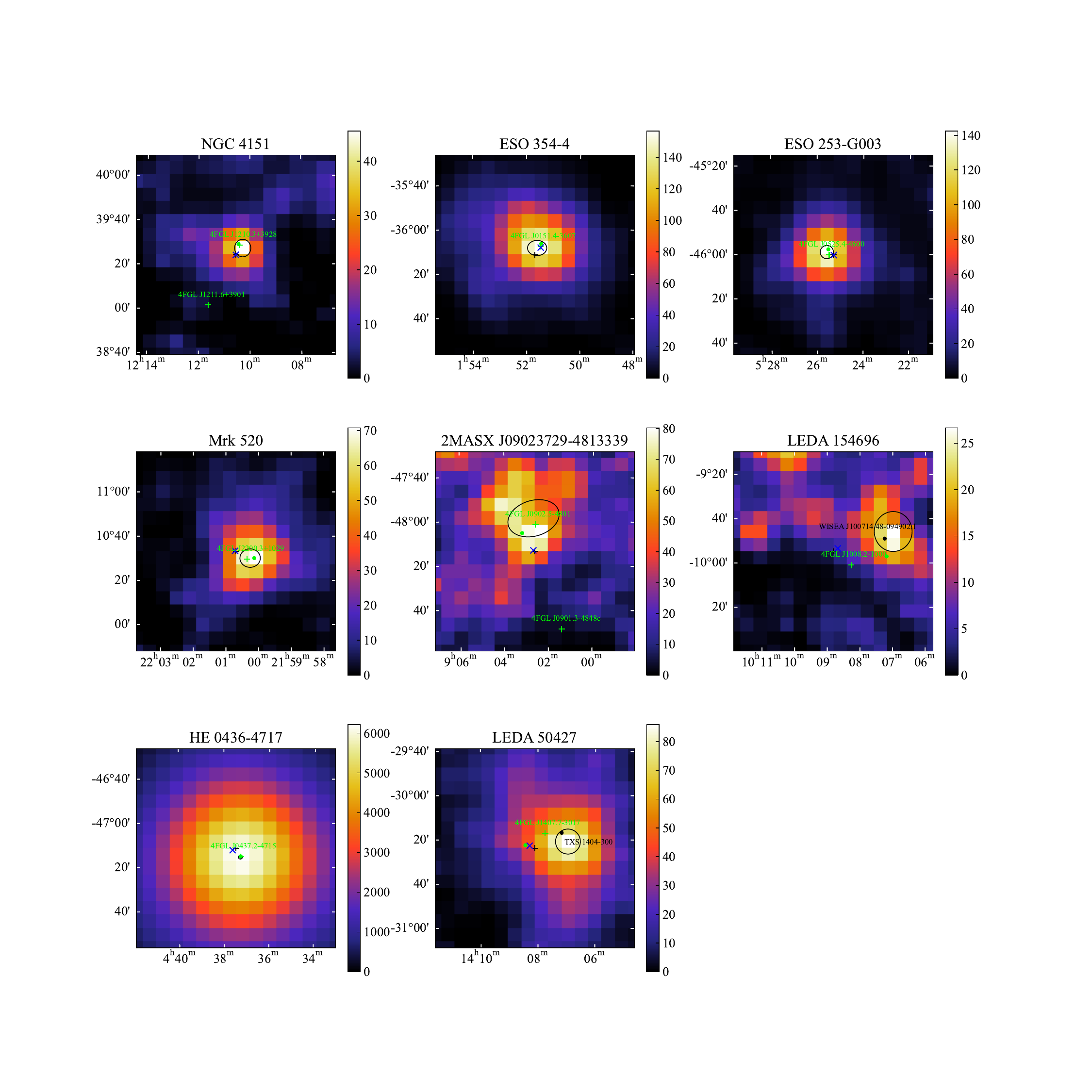}
\caption{\footnotesize
TS maps of the BASS-4FGL non-blazar sample.
The TS maps ($1.5^\circ \times 1.5^\circ,\, 0.1^\circ$/pixel) are derived in 0.1-300\,GeV for the 8 sources in the sample.
Black crosses indicate the infrared (WISE) position of the sources in BASS DR2 catalog\cite{Koss2022}.
The blue X represents the corresponding hard X-ray position\cite{Oh2018}.
Black ellipses indicate the 95$\%$ confidence localization error ellipses in this paper.
Green crosses and points indicate the best-fit positions and the counterparts from 4FGL-DR4, respectively.
For LEDA 154696 and LEDA50427, the black points indicate the possible spatial associations mentioned in this work.
The $x$ and $y$ axes are R.A. and decl. (J2000) in degrees.
}
\label{fig:TSmap_non-blazar}
\end{figure*}

\clearpage
\begin{table*}
\centering
\begin{threeparttable}
\footnotesize
\caption{\bf The $\gamma$-ray SED of the FGR sample}
\begin{tabular}{ccccc cccc}
\hline\hline
$E_{\rm min}$ & $E_{\rm max}$& $f_{\rm c}$ & TS & \\
\hline
 [MeV] & [MeV] &[$10^{-11}\rm ph\,s^{-1}\,cm^{-2}$]& \\
\hline
$3.0\times10^2$&$7.6\times10^2$&$<$6.72&0.6& \\
$7.6\times10^2$&$1.9\times10^3$&$1.66\pm 0.63$&7.17& \\
$1.9\times10^3$&$4.8\times10^3$&$0.47\pm 0.18$&6.97& \\
$4.8\times10^3$&$1.2\times10^4$&$0.11\pm 0.06$&4.09& \\
$1.2\times10^4$&$3.1\times10^4$&$0.05\pm 0.03$&5.25& \\
$3.1\times10^4$&$7.8\times10^4$&$<$0.03&0.4& \\
$7.8\times10^4$&$2.0\times10^5$&$0.02\pm 0.01$&8.6& \\
$2.0\times10^5$&$5.0\times10^5$&$<$0.01&0.0& \\
\hline
\end{tabular}
\begin{tablenotes}[flushleft]
\footnotesize 
\item {\rmfamily{\bf Notes.} $\gamma$-ray data point is calculated if TS $>=$4, and a 95\% confidence upper limit is given if TS $<$ 4.
All error bars represent 1 $\sigma$ uncertainty.}
\end{tablenotes}
\label{table:TS_SED}
\end{threeparttable}
\end{table*}

\clearpage
{\setlength{\tabcolsep}{4pt}
\begin{landscape}
\footnotesize
\begin{longtable}{lcccc ccccc ccccc cc}
\caption{\bf Properties of AGNs in the FGR sample}\\
\hline
\hline
Name & R.A. & Decl. & $D$ & $\log(\Mbh/\Msun)$ & $\log\nu L_{\rm 3\,GHz}$& Catalog & $\log\lamX$ &$\log\Lx$ &$\log\Rrx$ & $\log\Rrb$ & $\log \LGam$ & TS \\
& [$^\circ$] & [$^\circ$] & [Mpc]&&[erg s$^{-1}$]& & & [erg s$^{-1}$] & & & [erg s$^{-1}$] & \\
(1)&(2)&(3)&(4)&(5)&(6)&(7)&(8)&(9)&(10)&(11)&(12)&(13)&\\
\hline%
\endfirsthead%
\hline
\hline
Name & R.A. & Decl. & $D$ & $\log(\Mbh/\Msun)$ & $\log\nu L_{\rm 3\,GHz}$& Catalog & $\log\lamX$ &$\log\Lx$ &$\log\Rrx$ & $\log\Rrb$ & $\log \LGam$ & TS \\
(1)&(2)&(3)&(4)&(5)&(6)&(7)&(8)&(9)&(10)&(11)&(12)&(13)&\\
\hline
\endhead
\hline%
\endfoot%
NGC 424&17.8652&-38.0835&51.0&7.49&38.35&VLASS&-3.95&42.82&-4.48&0.59&$< 40.9$&4.64& \\
NGC 788&30.2769&-6.8159&59.1&8.18&37.5&VLASS&-4.51&43.51&-6.01&-0.9&$< 40.75$&0.0& \\
NGC 1194&45.9546&-1.1037&58.7&7.83&37.21&VLASS&-4.18&43.17&-5.96&-0.55&$< 41.0$&0.0& \\
LEDA 86269&71.0376&28.2169&46.1&7.98&38.01&VLASS&-3.29&43.1&-5.09&&$< 41.23$&6.22& \\
ESO 362-18&79.8992&-32.6578&53.9&7.11&37.79&VLASS&-2.13&43.23&-5.44&-0.41&$< 41.11$&6.48& \\
ESO 5-4&91.4219&-86.6316&28.2&7.40&38.17&SUMSS&-4.32&42.47&-4.29&0.35&$< 40.85$&0.0& \\
Mrk 1210&121.0244&5.1139&58.7&6.86&38.89&VLASS&-2.46&43.37&-4.48&0.83&$< 41.6$&5.64& \\
NGC 2788A&135.6644&-68.2268&57.7&8.26&&&-5.59&42.93&&&$< 41.04$&1.6& \\
MCG-5-23-16&146.9172&-30.9489&36.2&7.65&37.48&VLASS&-2.3&43.52&-6.03&-0.08&$< 40.7$&0.56& \\
NGC 3081&149.8731&-22.8263&32.5&7.67&37.05&VLASS&-4.66&43.01&-5.96&-0.53&$< 40.19$&0.0& \\
NGC 3227&155.8774&19.8651&22.9&6.77&37.87&VLASS&-2.22&42.85&-4.98&-0.14&$< 40.62$&1.75& \\
NGC 3393&162.0978&-25.1620&56.0&7.52&38.55&VLASS&-4.27&43.01&-4.46&0.65&$< 41.33$&11.03& \\
NGC 3516&166.6977&72.5687&38.9&7.39&37.22&VLASS&-2.7&43.31&-6.08&-1.05&$< 40.31$&0.0& \\
NGC 3783&174.7571&-37.7386&38.5&7.37&37.87&VLASS&-2.31&43.49&-5.62&-0.04&$< 41.46$&8.53& \\
NGC 4051&180.7900&44.5313&11.0&6.13&36.4&VLASS&-2.53&41.79&-5.39&-0.59&$< 39.35$&2.94& \\
NGC 4138&182.3742&43.6853&13.7&7.71&35.59&LOFAR&&41.74&-6.15&-1.78&$< 39.62$&0.95& \\
NGC 4235&184.2912&7.1916&26.6&7.28&37.1&VLASS&&42.51&-5.41&-0.21&$< 40.34$&0.0& \\
NGC 4253&184.6105&29.8129&55.9&6.82&38.39&VLASS&&42.99&-4.6&0.29&$< 40.92$&0.0& \\
NGC 4258&184.7396&47.3040&7.7&7.56&35.83&VLASS&-4.26&41.21&-5.38&-1.31&$< 39.1$&1.52& \\
NGC 4388&186.4449&12.6622&18.1&6.94&37.75&VLASS&-2.59&43.04&-5.29&0.78&$< 39.61$&3.74& \\
NGC 4395&186.4537&33.5469&4.8&5.45&34.31&LOFAR&-3.09&40.88&-6.57&-2.63&$< 39.01$&3.69& \\
NGC 4507&188.9026&-39.9093&50.5&7.81&38.08&VLASS&-3.31&43.75&-5.67&0.14&$< 40.64$&0.0& \\
NGC 4593&189.9143&-5.3442&37.2&6.88&37.09&VLASS&-2.06&43.16&-6.07&-0.81&$< 40.48$&0.46& \\
NGC 4941&196.0546&-5.5516&20.4&7.00&37.13&VLASS&-4.73&42.0&-4.87&-0.55&$< 39.71$&0.0& \\
NGC 4939&196.0597&-10.3395&42.1&7.75&37.17&VLASS&-4.41&42.75&-5.58&-0.49&$< 40.95$&5.36& \\
MCG-6-30-15&203.9740&-34.2956&30.4&6.60&36.9&VLASS&-1.74&42.82&-5.92&-0.58&$< 40.59$&3.58& \\
4U 1344-60&206.9000&-60.6177&55.5&9.09&&&-3.7&43.61&&&$< 42.03$&5.67& \\
NGC 5728&220.5995&-17.2530&37.5&8.25&37.48&VLASS&-4.95&43.15&-5.67&-0.27&$< 40.81$&0.0& \\
NGC 5899&228.7636&42.0498&45.1&7.96&37.65&VLASS&-3.96&42.7&-5.04&-0.32&$< 41.21$&3.66& \\
ESO 137-34&248.8083&-58.0800&34.1&7.48&&&-4.55&42.51&&&$< 40.51$&4.02& \\
NGC 6221&253.1930&-59.2170&11.9&6.72&&&-3.48&41.58&&&$< 40.13$&9.93& \\
NGC 6300&259.2481&-62.8206&13.2&6.77&37.43&SUMSS&-2.7&42.3&-4.87&&$< 40.16$&2.55& \\
ESO 103-35&279.5847&-65.4276&58.3&7.37&38.25&SUMSS&-1.85&43.64&-5.39&0.59&$< 41.32$&0.0& \\
Fairall 51&281.2250&-62.3647&60.0&7.11&37.93&SUMSS&-2.2&43.2&-5.26&-0.03&$< 41.3$&0.99& \\
NGC 6814&295.6691&-10.3236&22.8&7.04&36.83&VLASS&-2.63&42.6&-5.77&-0.09&$< 40.71$&6.77& \\
NGC 7172&330.5079&-31.8696&33.9&8.15&37.28&VLASS&-3.03&43.34&-6.06&-0.24&$< 41.05$&3.72& \\
NGC 7314&338.9425&-26.0504&16.8&6.30&36.38&VLASS&-2.05&42.29&-5.9&-0.83&$< 40.28$&0.0&
\label{table:faint source}
\end{longtable}
\begin{flushleft}
\footnotesize 
{\bf Notes.}\\ 
(1): The IR counterpart of the hard X-ray source in the BASS DR2 catalog\cite{Koss2022}.\\
(2) and (3): Right ascension and decl. (J2000) of the IR counterpart of the BAT AGN based on WISE positions\cite{Koss2022}.\\
(4): Distances from ref.\cite{Koss2022}.
Distances are redshift independent if $D<$ 60 Mpc.\\
(5): SMBH mass\cite{Koss2022}.\\
(6): Monochromatic radio luminosity at 3 GHz.\\
(7): Survey from which the radio luminosity is derived.\\
(8): Soft X-ray luminosity in unit of Eddington luminosity, mainly from the Swift X-Ray Telescope Point-source catalog (2SXPS)\cite{Evans2020}
and supplemented with the fourth XMM-Newton serendipitous source catalog\cite{Webb2020}.\\
(9): Hard X-ray luminosity from the 105-Month Swift-BAT All-sky Hard X-Ray Survey catalog\cite{Oh2018}.\\
(10): Hard X-ray radio-loudness parameter.\\
(11): The classical radio loudness parameter, where the flux of $B$ band is derived from the ref.\cite{Veron2010}.\\
(12): 95$\%$ confidence upper limit on the $\gamma$-ray luminosity integrated from 1\,GeV to 300\,GeV.\\
(13): $\rm TS$ value of the individual AGN.
\end{flushleft}
\end{landscape}
}

\clearpage
\setlength{\tabcolsep}{2pt}
\begin{table*}
\caption{\bf BASS-4FGL non-blazar sample}
\centering
\footnotesize 
\begin{threeparttable}
\begin{tabular}{lcccc cc}
\hline
\hline
Name &  4FGL NAME & ASSOC1& CLASS1& ASSOC2& CLASS2 \\
(1)&(2)&(3)&(4)&(5)&(6)\\
\hline
Circinus Galaxy&J1413.1-6519&Circinus galaxy&sey& \\
NGC 4151&J1210.3+3928&1E 1207.9+3945&bll& \\
ESO 354-4&J0151.4-3607&PMN J0151-3605&bcu& \\
ESO 253-G003&J0525.4-4600&PKS 0524-460&fsrq& \\
Mrk 520&J2200.3+1029&TXS 2157+102&bll& \\
2MASX J09023729-4813339&J0902.5-4801&PMN J0903-4805&bcu& \\
LEDA 154696&J1008.2-1000&&&CRATES J100710-095715&agn&\\
HE 0436-4717&J0437.2-4715&PSR J0437-4715&MSP& \\
LEDA 50427&J1407.7-3017&&&WISEA J140826.40-302231.5&unk& \\
\hline
\end{tabular}
\begin{flushleft}
\footnotesize 
{\bf Notes.}\\
(1): The {infrared (IR)} counterpart of the hard X-ray source in the BASS DR2 catalog\cite{Koss2022}.\\
(2): 4FGL-DR4 name of $\gamma$-ray source.\\
(3): {Name} of firmly-identified or associated counterparts of the $\gamma$-ray sources {from 4FGL-DR4}\cite{Ballet2023}.\\
(4) {and (6)}: Classifications of firmly-identified (given in uppercase letters) or associated (given in lowercase letters) counterparts of the $\gamma$-ray sources {from 4FGL-DR4\cite{Ballet2023}}.
The following abbreviations are used:
sey = Seyfert galaxy,
bll = BL Lacerate object,
bcu = blazar of unknown type,
agn = non-blazar active galaxy,
msp = millisecond pulsar,
fsrq = flat-spectrum radio quasar,
unk = unknown.\\
{(5): Name of low-confidence association or of enclosing extended source {from 4FGL-DR4}\cite{Ballet2023}.}
\end{flushleft}
\end{threeparttable}
\label{table:jet contamination}
\end{table*}

\clearpage
\begin{ThreePartTable}
\begin{TableNotes}[para,flushleft]
\footnotesize
{\bf Notes.}\\ 
(1): The IR counterpart of the hard X-ray source in the BASS DR2 catalog\cite{Koss2022}.\\
{(2): TS value.}\\
{(3)}: Nearby blazars of the FGR source\cite{Massaro2015}.\\
{(4)}: The separation between the FGR source and {each} nearby {blazar}.\\
{(5)}: The possible 4FGL counterpart of the blazar.\\
{(6)}: The separation between the blazar and its possible 4FGL counterpart.\\
{(7)}: The long radius of 95\% confidence localization error ellipse\cite{Ballet2023}.
\end{TableNotes}
\footnotesize
\begin{longtable}{lcccc cccc}
\caption{\bf FGR sources with nearby blazars}
\\
\hline
\hline
Name & {TS} & nearby blazar & $\delta_{\rm blaz}$ & 4FGL NAME & $\delta_{\rm 4FGL}$& Conf\_95\_SemiMajor \\
& && [$^\circ$]&& [$^\circ$]&[$^\circ$] & \\
(1)&(2)&(3)&(4)&(5)&(6)&(7)\\
\hline
\endfirsthead
\hline
\endfoot
\insertTableNotes
\endlastfoot
LEDA 86269&6.22&5BZB J0440+2750&0.82&J0440.8+2749&0.017&0.031& \\
NGC 1194&0.0&5BZB J0304-0054&0.28&J0304.5-0054&0.014&0.038& \\
NGC 3227&1.75&5BZQ J1024+1912&0.72&&&& \\
NGC 3393&11.03&5BZB J1046-2535&0.56&J1046.8-2534&0.024&0.045& \\
NGC 3516&0.0&5BZQ J1101+7225&0.40&&&& \\
&&5BZQ J1107+7232&0.07&&&& \\
NGC 4051&2.94&5BZB J1202+4444&0.28&J1202.4+4442&0.058&0.078& \\
&&5BZQ J1203+4510&0.65&&&& \\
NGC 4235&0.0&5BZG J1215+0732&0.60&J1215.1+0731&0.014&0.041& \\
NGC 4253&0.0&5BZB J1221+3010&0.73&J1221.3+3010&0.010&0.012& \\
&&5BZB J1217+3007&0.33&J1217.9+3007&0.008&0.010& \\
&&5BZQ J1217+2925&0.48&&&& \\
NGC 4258&1.52&5BZG J1221+4742&0.55&J1221.1+4742&0.009&0.047& \\
NGC 4939&5.36&5BZU J1303-1051&0.57&&&& \\
&&5BZQ J1305-1033&0.39&&&& \\
NGC 6814&6.77&5BZQ J1939-1002&0.73&&&& \\
\hline
\label{table:nearbyBlazar}
\end{longtable}
\end{ThreePartTable}

\clearpage
\begin{ThreePartTable}
\footnotesize
\begin{longtable}{lcccccc}
\caption{\bf 
Properties of nearby X-ray dim, radio-quiet, non-blazar AGN in the control sample}
\\
\hline
\hline
Name&R.A.&Decl.& $f_{\rm 3\,GHz}$& $\fx$\\
&[$^\circ$]&[$^\circ$]&[mJy]& $[10^{-12}\,\efluxunit]$ &\\
(1)&(2)&(3)&(4)&(5)&\\
\hline%
\endfirsthead
\hline
\hline
Name&R.A.&Decl.& $f_{\rm 3\,GHz}$& $\fx$\\
&[$^\circ$]&[$^\circ$]&[mJy]& $[10^{-12}\,\efluxunit]$ &\\
(1)&(2)&(3)&(4)&(5)&\\
\hline
\endhead
\hline%
\endfoot%
\hline%
\insertTableNotes
\endlastfoot
LEDA 136991&6.3850&68.3624&7.56&19.57&\\
IC 1657&18.5292&-32.6509&1.6&12.18&\\
NGC 454E&18.6039&-55.3970&&19.04&\\
NGC 678&27.3535&21.9974&1.7&6.63&\\
LEDA 137972&30.3848&68.4061&2.59&7.04&\\
LEDA 89913&30.5723&68.3627&28.62&14.77&\\
NGC 1125&42.9186&-16.6506&28.3&16.23&\\
MCG-1-9-45&52.8460&-5.1417&&6.94&\\
NGC 1566&65.0016&-54.9379&4.01&19.54&\\
UGC 3478&98.1966&63.6737&4.32&8.4&\\
NGC 2712&134.8770&44.9140&4.29&10.01&\\
Mrk 18&135.4934&60.1517&28.43&14.85&\\
IC 2461&139.9915&37.1910&2.22&19.11&\\
ESO 499-41&151.4807&-23.0569&3.88&18.06&\\
ESO 436-34&158.1856&-28.6102&2.56&16.55&\\
NGC 3718&173.1453&53.0680&11.63&12.24&\\
NGC 3786&174.9271&31.9094&6.96&14.63&\\
UGC 6732&176.3880&58.9782&3.24&14.69&\\
NGC 4180&183.2628&7.0389&6.5&17.58&\\
NGC 4500&187.8425&57.9646&8.89&5.39&\\
NGC 5033&198.3645&36.5939&5.78&6.26&\\
ESO 21-4&203.1693&-77.8446&&15.68&\\
ESO 383-18&203.3588&-34.0148&4.53&18.21&\\
NGC 5283&205.2740&67.6722&7.77&7.43&\\
NGC 5273&205.5349&35.6543&2.17&15.95&\\
ESO 138-1&252.8345&-59.2345&23.55&19.46&\\
NGC 7465&345.5040&15.9648&3.63&18.98&
\label{table:control sample}
\begin{TableNotes}[para,flushleft]
\footnotesize
{\bf Notes.}\\ 
(1): The IR counterpart of the hard X-ray source in the BASS DR2 catalog\cite{Koss2022}.\\
(2) and (3): Right ascension and decl. {(J2000)} of the IR counterpart of the BAT AGN based on WISE positions\cite{Koss2022}.\\
(4): 3 GHz radio flux (see \S\,\ref{sect:sample}).\\
{(5)}: The hard X-ray flux from the 105-month Swift-BAT All-sky Hard X-ray Survey catalog\cite{Oh2018}.
\end{TableNotes}
\end{longtable}
\end{ThreePartTable}

\clearpage
\noindent {\bf References}

\bibliographystyle{naturemag}
\bibliography{draft/corona}

\end{document}